\documentclass[12pt]{iopart}

\usepackage{amssymb,amsmath,amsfonts,amssymb}
\usepackage{graphics,graphicx,color}
\usepackage{numbysec}

\def\@abssec#1{\vspace{.05in}\footnotesize \parindent .2in
{\bf #1. }\ignorespaces}

\graphicspath{{/EPSF/}{Figures/}}
\DeclareGraphicsExtensions{.eps}

\newtheorem{theorem}{Theorem}[section]
\newtheorem{lemma}[theorem]{Lemma}

\newtheorem{corollary}[theorem]{Corollary}

\def \Rm {\mathbb R}

\def\arcsinh {\textrm{arcsinh}}
\def\ep{\varepsilon}

\def\C{\mathbb C}
\def\R{\mathbb R}
\def\S{\mathbb S}
\def\N{\mathbb N}

\def\pa{\partial}

\def\A{{\mathcal A}}

\newcommand{\eps}{\varepsilon}

\newcommand{\dint}{\displaystyle\int}
\newcommand{\pdr}[2]{\dfrac{\partial{#1}}{\partial{#2}}}
\newcommand{\pdrr}[2]{\dfrac{\partial^2{#1}}{\partial{#2}^2}}

\newcommand{\dr}[2]{\dfrac{d{#1}}{d{#2}}}

\newcommand{\cout}[1]{}

 \renewcommand{\arraystretch}{1.5}

\begin{document}
\numberbysection
\title{Inverse Transport Theory of Photoacoustics}

\author{Guillaume Bal}
\address{Department of Applied Physics \& Applied Mathematics,
  Columbia University, New York, NY 10027}
\ead{gb2030@columbia.edu}

\author{Alexandre Jollivet}
\address{Department of Applied Physics \& Applied Mathematics,
  Columbia University, New York, NY 10027}
\ead{aj2315@columbia.edu}

\author{Vincent Jugnon} \address{Centre de math{\'e}matiques
  appliqu{\'e}es, Ecole Polytechnique, 91128 Palaiseau, France and
  Department of Applied Physics \& Applied Mathematics, Columbia
  University, New York, NY 10027} \ead{jugnon@cmapx.polytechnique.fr}


\begin{abstract}
  We consider the reconstruction of optical parameters in a domain of
  interest from photoacoustic data. Photoacoustic tomography (PAT)
  radiates high frequency electromagnetic waves into the domain and
  measures acoustic signals emitted by the resulting thermal
  expansion.  Acoustic signals are then used to construct the
  deposited thermal energy map.  The latter depends on the
  constitutive optical parameters in a nontrivial manner. In this
  paper, we develop and use an inverse transport theory with internal
  measurements to extract information on the optical coefficients from
  knowledge of the deposited thermal energy map. We consider the
  multi-measurement setting in which many electromagnetic radiation
  patterns are used to probe the domain of interest. By developing an
  expansion of the measurement operator into singular components, we
  show that the spatial variations of the intrinsic attenuation and
  the scattering coefficients may be reconstructed. We also
  reconstruct coefficients describing anisotropic scattering of
  photons, such as the anisotropy coefficient $g(x)$ in a
  Henyey-Greenstein phase function model.  Finally, we derive
  stability estimates for the reconstructions.
\end{abstract}
 

\renewcommand{\thefootnote}{\fnsymbol{footnote}}
\renewcommand{\thefootnote}{\arabic{footnote}}

\renewcommand{\arraystretch}{1.1}





\section{Introduction}
\label{sec:intro}

Photoacoustic imaging is a recent medical imaging technique combining
the large contrast between healthy and unhealthy tissues of their
optical parameters with the high spatial resolution of acoustic
(ultrasonic) waves. Electromagnetic radiation, sent through a domain
of interest, generates some heating and a resulting thermal expansion
of the underlying tissues. The mechanical displacement of the tissues
generates acoustic waves, which then propagate through the medium and
are recorded by an array of detectors (ultrasound transducers).  The
photoacoustic effect is now being actively investigated for its
promising applications in medical imaging. We refer the reader to e.g.
\cite{KK-EJAM-08,PS-IP-07,XW-RSI-06} for recent reviews on the
practical and theoretical aspects of the method.

In an idealized setting revisited below, the electromagnetic source is
a very short pulse that propagates through the domain at a scale
faster than that of the acoustic waves. The measured acoustic signals
may then be seen as being emitted by unknown initial conditions.  A
first step in the inversion thus consists in reconstructing this
initial condition by solving an inverse source problem for a wave
equation. This inversion is relatively simple when the sound speed is
constant and full measurements are available. It becomes much more
challenging when only partial measurements are available and the sound
speed is not constant; see e.g.
\cite{ABJK-SR-09,CAB-IP-07,KK-EJAM-08,PS-IP-07,SU-IP-09}.

A second step consists of analyzing the initial condition
reconstructed in the first step and extracting information about the
optical coefficients of the domain of interest. The second step is
much less studied. The energy deposited by the radiation is given by
the product of $\sigma_a(x)$, the attenuation in the tissue and of
$I(x)$, the radiation intensity. The question is therefore what
information on the medium may be extracted from $\sigma_a I$. The
product can be plotted as a proxy for $\sigma_a$ when $I$ is more or
less uniform.  This, however, generates image distortions as has been
reported e.g.  in \cite{LPKW-PRE-08}. The extraction of e.g.
$\sigma_a$ from $\sigma_a I$ remains an essentially unsolved problem.

Two different regimes of radiation propagation should then be
considered. In thermoacoustic tomography (TAT), low frequency
(radio-frequency) waves with wavelengths much larger than the domain
of interest, are being used. We do not consider this modality here.
Rather, we assume that high frequency radiation is generated in the
near infra red (NIR) spectrum.  NIR photons have the advantage that
they propagate over fairly large distances before being absorbed.
Moreover, their absorption properties have a very large contrast
between healthy and cancerous tissues. In this regime, $I(x)$ may be
interpreted as a spatial density of photons propagating in the domain
of interest. The density of photons is then modeled by a transport
equation that accounts for photon propagation, absorption, and
scattering; see \eqref{eq:steadytr} below.

This paper concerns the reconstruction of the optical parameters in a
steady-state transport equation from knowledge of
$H(x):=\sigma_a(x)I(x)$. The derivation of the transport equation
given in \eqref{eq:steadytr} below is addressed in section
\ref{sec:main}.  We are concerned here with the setting of
measurements of $H(x)$ for different radiation patterns. Our most
general measurement operator $A$ is then the operator which to
arbitrary radiation patterns at the domain's boundary maps the
deposited energy $H(x)$. 

We analyze the reconstruction of the absorption and scattering
properties of the photons from knowledge of $A$. The main tool used in
the analysis is the decomposition of $A$ into singular components, in
a spirit very similar to what was done e.g. in
\cite{BJ-IP-09,BJ-IPI-08,choulli-stefanov-IP} (see also
\cite{B-IP-09}) in the presence of boundary measurements rather than
internal measurements. The most singular component is related to the
ballistic photons. We show that the analysis of that component allows
us to reconstruct the attenuation coefficient $\sigma_a$ and the
spatial component $\sigma_s$ of the scattering coefficient. The
anisotropic behavior of scattering is partially determined by the
second most singular term in $A$, which accounts for photons having
scattered only once in the domain. Although the full phase function of
the scattering coefficient cannot be reconstructed with the techniques
described in this paper, we show that the anisotropy coefficient
$g(x)$ that appears in the classical Henyey-Greenstein phase function
is uniquely determined by the measurements in spatial dimensions $n=2$
and $n=3$.  Moreover, all the parameters that can be reconstructed are
obtained with H\"older-type stability. We present the stability
results in detail.

When scattering is very large, then radiation is best modeled by a
diffusion equation characterized by two unknown coefficients, the
diffusion coefficient $D(x)$ and the attenuation coefficient
$\sigma_a(x)$.  This regime is briefly mentioned in section
\ref{sec:diff}.

The rest of the paper is structured as follows. Section \ref{sec:main}
is devoted to the derivation of the stationary inverse transport
problem starting from the transient equation for the short
electromagnetic pulse.  The main uniqueness and stability results are
also presented in detail in this section. The derivation of the
uniqueness and stability results is postponed to the technical sections
\ref{sec:forward} and \ref{sec:inverse}. The former section is devoted
to the decomposition of the albedo operator into singular components.
Useful results on the transport equation are also recalled. The latter
section presents in detail the proofs of the stability results
given in section \ref{sec:main}.

\section{Derivation and main results}
\label{sec:main}

\subsection{Transport and inverse wave problem}
\label{sec:wave}

The propagation of radiation is modeled by the following radiative
transfer equation
\begin{equation}
  \label{eq:transport}
  \begin{array}{ll}
  \dfrac1c \pdr{}t u(t,x,v) + T u(t,x,v) = S(t,x,v), & t\in\Rm, \,x\in\Rm^n,
  \,v\in \S^{n-1}.
  \end{array}
\end{equation}
We assume here that $S(t,x,v)$ is compactly supported in $t\geq0$ and
in $x$ outside of a bounded domain of interest $X$ we wish to probe.
The domain $X$ is assumed to be an open subset of $\R^n$ with $C^1$
boundary.  For the sake of simplicity, $X$ is also assumed to be
convex although all uniqueness and stability results in this paper
remain valid when $X$ is not convex.  Here, $c$ is light speed,
$\S^{n-1}$ is the unit sphere in $\Rm^n$ and $T$ is the transport
operator defined as
\begin{equation}
  \label{eq:T}
  Tu = v\cdot\nabla_x u + \sigma(x,v) u
   -\dint_{\S^{n-1}} k(x,v',v) u(t,x,v') dv',
\end{equation}
where $\sigma(x,v)$ is the total attenuation coefficient and
$k(x,v',v)$ is the scattering coefficient. Both coefficients are
assumed to be non-negative and bounded (by a constant $M<\infty$)
throughout the paper. We define
\begin{equation}
  \label{eq:sigma}
  \sigma_s(x,v) = \int_{\S^{n-1}} k(x,v,v') dv',\qquad
  \sigma_a(x,v) = \sigma(x,v) - \sigma_s(x,v).
\end{equation}
We also refer to $\sigma_s$ as a scattering coefficient and define
$\sigma_a$ as the intrinsic attenuation coefficient. We assume that
for (almost) all $(x,v)\in \Rm^n\times\S^{n-1}$, we have
$\sigma_a(x,v)\geq\sigma_0>0$. We also assume that $k(x,v,v')=0$ for
(almost) all $x\not\in X$.

The optical coefficients $k(x,v',v)$ and $\sigma(x,v)$ are the unknown
coefficients inside $X$ that we would like to reconstruct by probing
the domain $X$ by radiation modeled by $S(t,x,v)$. In photo-acoustics,
the emitted radiation generates some heating inside the domain $X$.
Heating then causes some dilation, which mechanically induces acoustic
waves. Such acoustic waves are measured at the boundary of the domain
$X$. After time reversion, the latter measurements allow us to infer
the intensity of the source of heating. This gives us internal
measurements of the solution $u$ of the transport equation
\eqref{eq:transport}. The objective of this paper is to understand
which parts of the optical parameters may be reconstructed from such
information and with which stability.

Before doing so, we need an accurate description of the propagation of
the acoustic waves generated by the radiative heating. The proper
model for the acoustic pressure is given by the following wave
equation (see e.g. \cite{FSS-PRE-07})
\begin{equation}
  \label{eq:wave}
  \square p(t,x) = \beta \pdr{}t H(t,x),
\end{equation}
where $\square$ is the d'Alembertian defined as 
\begin{equation}
  \label{eq:square}
  \square p = \dfrac{1}{c_s^2(x)} \pdrr pt - \Delta p,
\end{equation}
with $c_s$ the sound speed, where $\beta$ is a coupling coefficient
assumed to be constant and known, and where $H(t,x)$ is the thermal
energy deposited by the radiation given by
\begin{equation}
  \label{eq:H}
  H(t,x) = \dint_{\S^{n-1}} \sigma_a(x,v') u(t,x,v') dv'.
\end{equation}
Not surprisingly, the amount of heating generated by radiation is
proportional to the amount of radiation $u$ and to the rate of
(intrinsic) absorption $\sigma_a$.

As it stands, the problem of the reconstruction of the source term
$H(t,x)$ inside $X$ from measurements of $p(t,x)$ on the boundary
$\partial X$ is ill-posed, because $H$ is $(n+1)-$dimensional
whereas information on $(t,x)\in \Rm_+\times\partial X$ is
$n-$dimensional. What allows us to simplify the inverse acoustic
problem is the difference of time scales between the sound speed $c_s$
and the light speed $c$.

To simplify the analysis, we assume that $c_s$ is constant and rescale
time so that $c_s=1$. Then $c$ in \eqref{eq:transport} is replaced by
$\frac c{c_s}$, which in water is of order $2.3\,10^8/1.5\,10^3\approx
1.5\,10^5:=\frac1\eps\gg1$. The transport scale is therefore
considerably faster than the acoustic scale. As a consequence, when
the radiation source term $S(t,x,v)$ is supported on a scale much
faster than the acoustic scale, then $u$ is also supported on a scale
much faster than the acoustic scale and as a result $H(t,x)$ can be
approximated by a source term supported at $t=0$.

More specifically, let us assume that the source of radiation is defined
at the scale $\eps=\frac{c_s}c$ so that $S$ is replaced by
\begin{equation}
  \label{eq:Seps}
  S_\eps(t,x,v) = \dfrac1\eps \rho\big(\dfrac t\eps\big) S_0(x,v),
\end{equation}
where $\rho\geq0$ is a function compactly supported in
$t\in(0,\infty)$ such that $\int_{\Rm_+}\rho(t)dt=1$. The transport
solution then solves
\begin{displaymath}
  \dfrac1c \pdr{}t u_\eps(t,x,v) + T u_\eps(t,x,v) = S_\eps(t,x,v).
\end{displaymath}
With $c_s=1$, we verify that $u_\eps$ is given by
\begin{equation}
  \label{eq:ueps}
  u_\eps(t,x,v) = \dfrac{1}{\eps} u\big(\frac t\eps,x,v\big),
\end{equation}
where $u$ solves \eqref{eq:transport} with $S(t,x,v)=\rho(t)S_0(x,v)$.
Because $\sigma_a\geq\sigma_0>0$, we verify that $u$ decays
exponentially in time. This shows that $u_\eps$ lives at the time
scale $\eps$ so that $H_\eps(t,x)=\int_{\S^{n-1}} \sigma_a(x,v)
u_\eps(t,x,v)dv$ is also primarily supported in the vicinity of $t=0$.

Let us formally derive the equation satisfied by $p_\eps$ when
$\eps\to0$. Let $\varphi(t,x)$ be a test function and define
$(\cdot,\cdot)$ as the standard inner product on $\Rm\times\Rm^n$.
Then we find that 
\begin{displaymath}
  (\square p_\eps,\varphi) = \beta(\pdr {H_\eps}t,\varphi)
  =-\beta(H_\eps,\pdr \varphi t) = -\beta(H,\pdr{\varphi}t (\eps t))),
\end{displaymath}
where $H(t,x)=\int_{\S^{n-1}} \sigma_a(x,v)u(t,x,v)dv$ with $u$
defined in \eqref{eq:ueps} and where we have used the change of
variables $t\to\eps t$.  The latter term is therefore equal to
\begin{displaymath}
  -\beta(H,\pdr{\varphi}t(0)) = -\beta \pdr{\varphi}t(0)
  \dint_{\Rm\times \S^{n-1}} \sigma_a(x,v) u(t,x,v) dt dv,
\end{displaymath}
up to a small term for $\varphi$ sufficiently smooth. We thus find that
$p_\eps$ converges weakly as $\eps\to0$ to the solution $p$ of
the following wave equation
\begin{equation}
  \label{eq:limitwave}
  \begin{array}{ll}
  \square p =0 \quad &t>0, \quad x\in \Rm^n \\
  p(0,x) = H_0(x) :=
  \dint_{\Rm\times \S^{n-1}} \hspace{-.7cm}\sigma_a(x,v) u(t,x,v) dt dv,
   \quad &x\in\Rm^n \\
  \pdr pt(0,x) =0 \quad &x\in\Rm^n.
  \end{array}
\end{equation}

The inverse problem for the wave equation is now well-posed. The
objective is to reconstruct $H_0(x)$ for $x\in X$ from measurements of
$p(t,x)$ for $t\geq0$ and $x\in\partial X$. Such an inverse problem
has been extensively studied in the literature. We refer the reader to
e.g. \cite{SU-IP-09} for a recent inversion with (known) variable
sound speed.

In this paper, we assume that $H_0(x)$ has been reconstructed
accurately as a functional of the radiation source $S_0(x,v)$.  Our
objective is to understand which parts of the optical parameters
$\sigma(x,v)$ and $k(x,v',v)$ can be reconstructed from knowledge of
$H_0(x)$ for a given set of radiations $S_0(x,v)$.  Note that we will
allow ourselves to generate many such $S_0(x,v)$ and thus consider a
multi-measurement setting.

The time average $u(x,v):=\int u(t,x,v)dt$ satisfies a
closed-form steady state transport equation given by
\begin{equation}
  \label{eq:steadytr0}
  T u = S_0(x,v), \qquad (x,v)\in\Rm^n\times\S^{n-1},
\end{equation}
as can be seen by averaging \eqref{eq:transport} in time since
$\int \rho(t)dt=1$. 

\subsection{Inverse transport with internal measurements}
\label{sec:internal}

Since $S_0(x,v)$ is assumed to be supported
outside of the domain $X$ and scattering $k(x,v',v)=0$ outside
of $X$, the above transport equation may be replaced by a boundary
value problem of the form
\begin{equation}
  \label{eq:steadytr}
  \begin{array}{ll}
  v\cdot\nabla_x u + \sigma(x,v) u - \dint_{\S^{n-1}}
   k(x,v',v) u(x,v') dv' =0, \quad & (x,v)\in X\times \S^{n-1} \\
  u(x,v) = \phi(x,v) \quad & (x,v) \in \Gamma_-, 
  \end{array}
\end{equation}
where the sets of outgoing and incoming boundary radiations are given by
\begin{equation}
  \label{eq:Gammapm}
  \Gamma_\pm = \{(x,v)\in \partial X\times \S^{n-1},\quad
   \pm v\cdot \nu(x) >0 \},
\end{equation}
where $\nu(x)$ is the outward normal to $X$ at $x\in\partial X$ and
$\phi(x,v)$ is an appropriate set of incoming radiation conditions obtained
by solving $v\cdot\nabla_x u+\sigma_a(x,v)u=S_0$ outside of $X$ assuming
that $\sigma_a(x,v)$ is known outside of $X$.

We are now ready to state the inverse transport problem of interest in
this paper. It is well known that \eqref{eq:steadytr} admits a unique
solution in $L^1(X\times \S^{n-1})$ when $\phi(x,v)\in
L^1(\Gamma_-,d\xi)$, where $d\xi=|v\cdot\nu(x)|d\mu(x)dv$ with $d\mu$
the surface measure on $\partial X$. We thus define the albedo
operator as
\begin{equation}
  \label{eq:A}
  \begin{array}{lrcl}
   A : & L^1(\Gamma_-,d\xi) & \to &  L^1(X) \\
   & \phi(x,v) &\mapsto & A\phi(x) = 
  H(x) := \dint_{\S^{n-1}} \sigma_a(x,v) u(x,v) dv.
  \end{array}
\end{equation}
The inverse transport problem with angularly averaged internal
measurements thus consists of understanding what can be reconstructed
from the optical parameters $\sigma(x,v)$ and $k(x,v',v)$ from
complete or partial knowledge of the albedo operator $A$. We also
wish to understand the stability of such reconstructions.

\subsection{Albedo operator and decomposition}
\label{sec:albedo}

The inverse transport problem and its stability properties are solved
by looking at a decomposition of the albedo operator into singular 
components. Let $\alpha(x,x',v')$ be the Schwartz kernel of the 
albedo operator $A$, i.e., the distribution such that 
\begin{equation}  \label{eq:Schkernel}
  A\phi(x) = \dint_{\Gamma_-} \alpha(x,x',v') \phi(x',v') d\mu(x') dv'.
\end{equation}
The kernel $\alpha(x,x',v')$ corresponds to measurements of $H(x)$ at
$x\in X$ for a radiation condition concentrated at $x'\in\partial X$
and propagating with direction $v'\in \S^{n-1}$. Such a kernel can
thus be obtained as a limit of physical experiments with sources
concentrated in the vicinity of $(x',v')$ and detectors concentrated
in the vicinity of $x$.

The kernel $\alpha(x,x',v')$ accounts for radiation propagation inside
$X$, including all orders of scattering of the radiation with the
underlying structure. It turns out that we can extract from
$\alpha(x,x',v')$ singular components that are not affected by
multiple scattering. Such singular components provide useful
information on the optical coefficients. Let us define 
the ballistic part of transport as the solution of
\begin{equation}
  \label{eq:ballistic}
  v\cdot\nabla_x u_0 + \sigma(x,v) u_0 =0, 
  \quad \mbox{ in } X\times \S^{n-1}, \qquad 
  u_0=\phi, \quad \mbox{ on } \Gamma_-. 
\end{equation}
Then for $m\geq1$, we define iteratively
\begin{equation}
  \label{eq:multscat}
  v\cdot\nabla_x u_m + \sigma(x,v) u_m =\dint  k(x,v',v) u_{m-1}(x,v') dv',
  \,\,\mbox{ in } X\times \S^{n-1}, \quad 
  u_m=0, \,\, \mbox{ on } \Gamma_-. 
\end{equation}

This allows us to decompose the albedo operator as
\begin{equation}
  \label{eq:decA}
  A = A_0 + A_1 + {\mathcal G}_2,
\end{equation}
where $A_{k}$ for $k=0,1$ are defined as $A$ in \eqref{eq:A} with $u$
replaced by $u_k$ and where ${\mathcal G}_2$ is defined as
$A-A_0-A_1$. Thus, $A_0$ is the contribution in $A$ of particles that
have not scattered at all with the underlying structure while $A_1$ is
the contribution of particles that have scattered exactly once.

Let $\alpha_{k}$ for $k=0,1$ be the Schwartz kernel of $A_k$ and
$\Gamma_2$ the Schwartz kernel of ${\mathcal G}_2$ using the same
convention as in \eqref{eq:Schkernel}.  We define $\tau_\pm(x,v)$ for
$x\in X$ and $v\in \S^{n-1}$ as $\tau_\pm(x,v)=\inf\{s\in\Rm_+ | x\pm
sv\not\in X\}$. Thus, $\tau_\pm(x,v)$ indicates the time of escape from
$X$ of a particle at $x$ moving in direction $\pm v$. On $\partial X$,
we also define $\delta_{\{x\}}(y)$ as the distribution such that
$\int_{\partial X} \delta_{\{x\}}(y)\phi(y)d\mu(y)=\phi(x)$ for any
continuous function $\phi$ on $\partial X$. Finally, we define 
the following terms that quantify attenuation. We define the function
$E(x_0,x_1)$ on $X\times \partial X$ as 
\begin{equation}
  \label{eq:E}
  E(x_0,x_1) = \exp \Big(-\dint_0^{|x_0-x_1|}
   \sigma\big(x_0-s\frac{x_0-x_1}{|x_0-x_1|},\frac{x_0-x_1}{|x_0-x_1|}\big)ds
   \Big).
\end{equation}
We still denote by $E$ the function defined above for $x_1$ in $X$.
Then by induction on $m$, we define
\begin{equation}
  \label{eq:Ek}
  E(x_1,\ldots,x_m) = E(x_1,\ldots,x_{m-1})E(x_{m-1},x_m).
\end{equation}
The latter term measures the attenuation along the broken path
$[x_1,\ldots,x_m]$.

Then we have the following result.
\begin{theorem}
  \label{thm:decomp}
  Let $\alpha_0$, $\alpha_1$, and $\Gamma_2$ be the Schwartz kernels
  defined as above. Then we have:
  \begin{equation}
    \label{eq:alpha01}
    \begin{array}{rcl}
 \alpha_0(x,x',v') &=& \sigma_a(x,v')
   \exp \Big(-\dint_0^{\tau_-(x,v')}\sigma(x-sv',v') ds\Big)
   \delta_{\{x-\tau_-(x,v')v'\}}(x') \\
  \alpha_1(x,x',v') &=& |\nu(x')\cdot v'| \\
   && \dint_0^{\tau_+(x',v')} \hspace{-.5cm} \sigma_a(x,v)
   \dfrac{E(x,x'+t'v',x')}{|x-x'-t'v'|^{n-1}}
   k(x'+t'v',v',v)
  |_{v=\frac{x-x'-t'v'}{|x-x'-t'v'|}} dt'.  
    \end{array}
  \end{equation}
  Moreover, we have the bound
  \begin{equation}
    \label{eq:Gamma2}
    \begin{array}{rcl}
   \dfrac{\Gamma_2(x,x',v')}{|\nu(x')\cdot v'|} \in L^\infty(X\times\Gamma_-)
   &\mbox{ when } & n=2 \\
  \dfrac{|x-x'-((x-x')\cdot v')v'|^{n-3}\Gamma_2(x,x',v')}
   {|\nu(x')\cdot v'|} \in L^\infty(X\times\Gamma_-)
   &\mbox{ when } & n\geq3.
    \end{array}
  \end{equation}
\end{theorem}
This theorem will be proved in section \ref{sec:forward}. The results
show that $\alpha_0$ is more singular than $\alpha_1$ and $\Gamma_2$.
It turns out that $\alpha_1$ is also more singular than $\Gamma_2$ in
the sense that it is asymptotically much larger than $\Gamma_2$ in the
vicinity of the support of the ballistic part $\alpha_0$. That this is
the case is the object of the following result (see also Lemma
\ref{lem:norm_scat} \eqref{eq:norm_simple} and \eqref{eq:norm_alpham}
for ``$m=1$'').  For any topological space $Y$, we denote by
${\mathcal C}_b(Y)$ the set of the bounded continuous functions from
$Y$ to $\R$.
\begin{theorem}
  \label{thm:asympalpha1}
  Let us assume that $\sigma\in {\mathcal C}_b(X\times\S^{n-1})$ and
  that $k\in{\mathcal C}_b(X\times\S^{n-1}\times\S^{n-1})$.  For
  $(x',v')\in\Gamma_-$, let $x=x'+t_0'v'$ for some $t_0'\in
  (0,\tau_+(x',v'))$. Let $v'^\perp\in \S^{n-1}$ be such that  
  $v'\cdot v'^\perp=0$. Then we have the following asymptotic expansion:
  \begin{equation}
    \label{eq:asyalpha1}
     \begin{array}{ll}
    \dfrac{\alpha_1(x+\eps v'^\perp,x',v')}{E(x,x')|\nu(x')\cdot v'|}
    =\Big(\ln\dfrac1\eps \Big)
  \big(  \chi(x,v',v')+\chi(x,v',-v')\big) 
  + o\Big(\ln\dfrac1\eps\Big) \\
    \dfrac{\alpha_1(x+\eps v'^\perp,x',v')}{E(x,x')|\nu(x')\cdot v'|}
    =\dfrac{1}{\eps^{n-2}}\dint_0^\pi \sin^{n-3}\theta
     \chi(x,v',v(\theta)) d\theta + o\Big(\dfrac{1}{\eps^{n-2}}\Big),
     \end{array}
  \end{equation}
  for $n=2$ and $n\geq3$, respectively, where we have defined the
  functions $\chi(x,v',v)=\sigma_a(x,v) k(x,v',v)$ and
  $v(\theta)=\cos\theta v' + \sin\theta v'^\perp$.
\end{theorem}
Theorem \ref{thm:asympalpha1} will be proved in section \ref{sec:forward}.

We thus observe that $\alpha_1$ blows up a priori faster than $\Gamma_2$ as
$\eps\to0$, i.e., as the observation point $x$ becomes closer to the
segment where the ballistic term $\alpha_0$ is supported.  This
singularity allows us to obtain information on the optical
coefficients that is not contained in the ballistic part $\alpha_0$.
Moreover, because of the singular behavior of $\alpha_1$, such
information can be reconstructed in a stable manner.

\subsection{Stability estimates}
\label{sec:stab}

As mentioned above, the singular behaviors of $\alpha_0$ and $\alpha_1$
allow us to extract them from the full measurements $\alpha$.
Moreover, such an extraction can be carried out in a stable fashion,
in the sense that small errors in the measurement of the albedo
operator translates into small errors in the extraction of the terms
characterizing $\alpha_0$ and $\alpha_1$.

More precisely, let $A$ be the albedo operator corresponding to
optical parameters $(\sigma,k)$ and $\tilde A$ the operator
corresponding to the optical parameters $(\tilde\sigma,\tilde k)$. From
now on, a term superimposed with the $\tilde{}$ sign means a term
calculated using the optical parameters $(\tilde\sigma,\tilde k)$
instead of $(\sigma,k)$.  For instance $\tilde E(x,y)$ is the
equivalent of $E(x,y)$ defined in \eqref{eq:E} with $(\sigma,k)$
replaced by $(\tilde\sigma,\tilde k)$. 

We first derive the stability of useful functionals of the
optical parameters in terms of errors made on the measurements.  Let
us assume that $A$ is the ``real'' albedo operator and that $\tilde A$
is the ``measured'' operator. We want to obtain error estimates on the
useful functionals of the optical parameters in terms of appropriate
metrics for $A-\tilde A$. We obtain the following two results.
The first result pertains to the stability of the ballistic term
in the albedo operator:
\begin{theorem}
  \label{thm:ballistic}
  Let $A$ and $\tilde A$ be two albedo operators and $(x',v')\in\Gamma_-$.
  Then we obtain that
  \begin{equation}
    \label{eq:stabballistic}
    \begin{array}{ll}
    &\dint_0^{\tau_+(x',v')} \Big| \sigma_a(x'+tv',v')
     e^{-\int_0^t \sigma(x'+sv',v')ds }-
   \tilde\sigma_a(x'+tv',v')
     e^{-\int_0^t \tilde\sigma(x'+sv',v')ds } \Big| dt\\
   \leq &\|A-\tilde A\|_{{\mathcal L}(L^1(\Gamma_-,d\xi);L^1(X))}.
    \end{array}
  \end{equation}
\end{theorem}
Theorem \ref{thm:ballistic} is proved in section \ref{sec:ballistic}.

The stability result obtained from the single scattering component 
is based on a singular behavior obtained in the vicinity of the ballistic
component. Such a behavior cannot be captured by the $L^1$ norm used above.
Instead, we define $\Gamma_1=\alpha-\alpha_0$ as the Schwartz kernel of
the albedo operator where the ballistic part has been removed, i.e., for
measurements that are performed away from the support of the ballistic
part. Our stability results are obtained in terms of errors on $\Gamma_1$
rather than on $A$. We can then show the following stability result.
\begin{theorem}
  \label{thm:stabscat}
  Let us assume that $(\sigma,\tilde \sigma)\in {\mathcal
    C}_b(X\times\S^{n-1})^2$ and that $(k,\tilde k)\in{\mathcal
    C}_b(X\times\S^{n-1}\times\S^{n-1})^2$.  Let $(x,x')\in
  X\times\partial X$ and define $v'=\frac{x-x'}{|x-x'|}$.  Let
  $v'^\perp\in \S^{n-1}$ such that $v'\cdot v'^\perp=0$. In dimension
  $n=2$, we have
  \begin{equation}
    \label{eq:dim2}
    \begin{array}{ll}
     &\Big| E(x,x') (\chi(x,v',v')+\chi(x,v',-v')) -
     \tilde E(x,x') (\tilde \chi(x,v',v')+\tilde \chi(x,v',-v')) \Big|\\
   \leq & \Big\| \dfrac{(\Gamma_1-\tilde\Gamma_1)(x,x',v')}
   {|\nu(x')\cdot v'| w_2(x,x',v')}
    \Big\|_{L^\infty(X\times \Gamma_-)},
    \end{array}
  \end{equation}
  where
  $w_2(x,x',v')=1+\ln\big(\frac{|x-x'-\tau_+(x',v')v'|-(x-x'-\tau_+(x',v')v')
    \cdot v'}{|x-x'|-(x-x')\cdot v'}\big)$. When $n\geq3$, we have
  \begin{equation}
    \label{eq:dim3}
    \begin{array}{ll}
      \Big|\dint_0^\pi \sin^{n-3}(\theta) \Big(E(x,x')\chi(x,v',v(\theta))
     -\tilde E(x,x')\tilde \chi(x,v',v(\theta))\Big) d\theta \Big| \\
     \leq \Big\|\dfrac{(\Gamma_1-\tilde\Gamma_1)(x,x',v')}
     {|\nu(x')\cdot v'|w_n(x,x',v')}
    \Big\|_{L^\infty(X\times \Gamma_-)},
    \end{array}
  \end{equation}
  where $w_n(x,x',v')=|x-x'-((x-x')\cdot v') v'|^{2-n}$ and where we
  use the same notation as in Theorem \ref{thm:asympalpha1}.
\end{theorem}
Theorem \ref{thm:stabscat} is proved in section \ref{sec:single}.

Such results do not grant uniqueness of the reconstruction of the
optical parameters in the most general setting. However, they do
provide stable, unique, reconstructions in several settings of
interest.

\subsection{Scattering-free setting}
\label{sec:scatteringfree}

Let us first assume that $k\equiv0$ so that $\sigma\equiv\sigma_a$.
Then knowledge of the albedo operator uniquely determines
$\sigma_a(x,v)$ for all $x\in X$ and $v\in \S^{n-1}$. 

Indeed, we deduce from Theorem \ref{thm:ballistic} that 
\begin{displaymath}
  \sigma_a(x'+tv',v')  e^{-\int_0^t \sigma_a(x'+sv',v')ds }
 = -\dr{}t \Big(e^{-\int_0^t \sigma_a(x'+sv',v')ds }\Big)
\end{displaymath}
is uniquely determined and hence $e^{-\int_0^t \sigma_a(x'+sv',v')ds
}$ since the latter equals $1$ when $t=0$. Taking the derivative of the
negative of the logarithm of the latter expression gives us
$\sigma(x'+tv',v')$ for all $(x',v')\in\Gamma_-$ and $t>0$ and hence
$\sigma(x,v)$ for all $(x,v)\in X\times \S^{n-1}$.

Moreover, we have the following stability result.
\begin{theorem}
  \label{thm:stabscatfree}
  Recalling
  that $\sigma_a(x,v)$ is bounded from above and below by positive
  constants, we find that when $k\equiv0$,
  \begin{equation}
    \label{eq:stabscatfree}
    \|\sigma_a-\tilde\sigma_a\|
  _{L^\infty(\S^{n-1};L^1(X))}
    \leq C \|A-\tilde A\|_{{\mathcal L}(L^1(\Gamma_-,d\xi);L^1(X))}.
  \end{equation}
  Here, $C$ is a constant that depends on the uniform bound $M$.
\end{theorem}
The above theorem is proved in section \ref{sec:ballistic}.  Note that
the above result is local in $x'$ and $v'$. In other words,
$\sigma_a(x,v')$ is uniquely determined by
$\{\alpha(y,x-\tau_-(x,v')v',v')|y=x+tv'\mbox{ for }
-\tau_-(x,v')<t<\tau_+(x,v')\}$, that is by the experiment that
consists of sending a beam of radiation in direction $v'$ passing
through the point $x$ (at least asymptotically since such a transport
solution is not an element in $L^1(X\times\S^{n-1})$).

\subsection{Reconstruction of the spatial optical parameters}
\label{sec:rec}

We now assume that $k\not=0$. Then the ballistic component of $A$ and
the estimate in Theorem \ref{thm:ballistic} allow us to uniquely reconstruct
both $\sigma_a(x,v)$ and $\sigma(x,v)$ under the assumption that
\begin{equation}
  \label{eq:symassump}
  \sigma_a(x,v)=\sigma_a(x,-v),\qquad
  \sigma(x,v) = \sigma(x,-v).
\end{equation}

We recall that $\sigma_a(x,v)$ is bounded from below by $\sigma_0>0$.
For technical reasons, we also assume that $\sigma(x,v)$ is known in
the $\delta_0-$vicinity of $\partial X$, i.e., for all
$(x,v)\in\Rm^n\times\S^{n-1}$ such that ${\rm
  dist}(x,\partial X)<\delta_0$ for some $\delta_0>0$. Such a
hypothesis is not very restrictive from a practical viewpoint.

We denote by $W^{-1,1}(X)$ the Banach space of the continuous linear functionals on the Banach space 
$W_0^{1,\infty}(X):=\{\phi\in L^\infty(X)\ | \ {\rm supp}\phi\subset X,\ \nabla \phi\in L^\infty(X,\C^{n})\}$ 
(where the gradient $\nabla$ is understood in the distributional sense).

Under the above assumptions, we have the following result.
\begin{theorem}
  \label{thm:stabsym} 
  The coefficients $\sigma_a(x,v)$ and $\sigma(x,v)$ are uniquely
  determined by the albedo operator $A$. Moreover, we have the 
  following stability estimate
  \begin{equation}
    \label{eq:stabsym}
    \begin{array}{ ll}
     \,\,
   \|\sigma-\tilde\sigma\|_{L^\infty(\S^{n-1};W^{-1,1}(X))}
    \\+\, 
  \|\sigma_a-\tilde\sigma_a\|_{L^\infty(\S^{n-1};L^1(X))}
      \,\leq \,C \|A-\tilde A\|_{{\mathcal L}(L^1(\Gamma_-,d\xi);L^1(X))}.
    \end{array}
  \end{equation}
  The constant $C$ depends on the parameter $\delta_0$ as well as the
  uniform bounds $\sigma_0$ and $M$.
\end{theorem}
This theorem is proved in section \ref{sec:ballistic}.  The above
result is also local in $x'$ and $v'$. In other words,
$\sigma_a(x,v')$ and $\sigma(x,v')$ are uniquely determined by
$\{\alpha(y,x-\tau_-(x,v')v',v')|y=x+tv'\mbox{ for }
-\tau_-(x,v')<t<\tau_+(x,v')\}$, i.e., by the experiment that consists
of sending a beam of radiation in direction $v'$ passing through the
point $x$.

Stability estimates may be obtained in stronger norms for $\sigma$
provided that a priori regularity assumptions be imposed. We show the
\begin{corollary}
  \label{cor:regsigma} Let us assume that $\sigma$ and $\tilde\sigma$ 
  are bounded in $L^\infty(\S^{n-1},W^{r,p}(X))$ by $C_0$ for $p>1$ and $r>-1$. Then for
  all $-1\leq s\leq r$, we have
  \begin{equation}
    \label{eq:postbdsigma}
    \|\sigma-\tilde\sigma\|_{L^\infty(\S^{n-1};W^{s,p}(X))}
    \leq C \|A-\tilde A\|^{\frac 1p\frac{r-s}{1+r}}
   _{{\mathcal L}(L^1(\Gamma_-,d\xi);L^1(X))},
  \end{equation}
where the constant $C$ depends on $C_0$ and on the uniform bounds $\sigma_0$ and $M$.
\end{corollary}
The corollary is proved in section \ref{sec:ballistic}.

\subsection{Application to Henyey-Greenstein kernels}
\label{sec:HG}
Let us assume that $\sigma\in {\mathcal C}_b(X\times\S^{n-1})$ and
  that $k\in{\mathcal C}_b(X\times\S^{n-1}\times\S^{n-1})$ and let us assume again that $\sigma(x)$ is known in
the $\delta_0-$vicinity of $\partial X$, i.e., for all
$x\in\Rm^n$ such that ${\rm
  dist}(x,\partial X)<\delta_0$ for some $\delta_0>0$ .

The stability estimate \eqref{eq:stabballistic} allows one to uniquely
reconstruct $\sigma_a$ and $\sigma$ under the symmetry hypothesis
\eqref{eq:symassump}, which is quite general physically.  Indeed, even
when attenuation is anisotropic, there is no reason to observe
different attenuations in direction $v$ and direction $-v$.
The stability estimate in Theorem \ref{thm:stabscat} provides additional
information on the optical coefficients, but not enough to fully
reconstruct the scattering kernel $k(x,v',v)$. 

In dimension $n=2$,
we gain information only on $k(x,v',v')+k(x,v',-v')$.
In dimension $n\geq3$, we garner information about 
$\int_0^\pi \chi(x,v',\cos\theta v'+\sin\theta v'^\perp)d\theta$
for all $v'^\perp$ orthogonal to $v'$. The integration in $\theta$
means that one dimension of information is lost in the measurements.
Thus, $3n-3$ dimensions of information are available on the 
$(3n-2)-$dimensional object $k(x,v',v)$.

Let us consider the case of an isotropic absorption
$\sigma_a=\sigma_a(x)$ and isotropic scattering in the sense that
$k(x,v',v)=k(x,v'\cdot v)$. In such a setting, $k(x,v'\cdot v)$
becomes $(n+1)-$dimensional. Yet, available data in dimension $n\ge 2$
give us information on
\begin{eqnarray}
\sigma_g(x)&:=&k(x,-1)+k(x,1), \textrm{ when }n=2,\label{g1}\\
\sigma_g(x)&:=& \dint_0^\pi k(x,\cos\theta)\sin^{n-3}\theta d\theta, \textrm{ when }n\ge 3.\label{g2}
\end{eqnarray} 
This is different information from the normalization in
\eqref{eq:sigma}
\begin{displaymath}
  \sigma_s(x) = \dint_{\S^{n-1}} k(x,v'\cdot v) dv'
  = |\S^{n-2}| \dint_0^\pi k(x,\cos\theta) \sin^{n-2}\theta d\theta.
\end{displaymath}
As a consequence, if $k(x,\cos\theta)$ is of the form
$\sigma_s(x)f(x,\cos\theta)$, where $f(x,\cos\theta)$ is parameterized
by one function $g(x)$, then we have a chance of reconstructing $g(x)$
from knowledge of $\sigma_g(x)$ and $\sigma_s(x)$ provided $\sigma_s(x)>0$ (where $\sigma_s=\sigma-\sigma_a$). 

This occurs for the classical Henyey-Greenstein (HG) phase function in
dimensions $n=2$ and $n=3$, where
\begin{eqnarray}
k(x,\lambda):=\sigma_s(x){1-g^2(x)\over 2\pi(1+g(x)^2-2g(x)\lambda)},\textrm{ when }n=2,\label{g3}\\
k(x,\lambda):=\sigma_s(x){1-g^2(x)\over 4\pi(1+g(x)^2-2g(x)\lambda)^{3\over 2}},\textrm{ when }n\ge 3,\label{g4}
\end{eqnarray} 
where $g\in C_b(X)$ and $0
\le g(x)<1$ for a.e. $x\in X$. Note that 
\begin{equation}
\sigma_g(x)=\sigma_s(x)h(g(x)),\label{g5}
\end{equation} 
for $x\in X$ where the function $h:[0,1)\to \R$ is given by 
\begin{eqnarray}
h(\kappa):={1+\kappa^2\over \pi(1-\kappa^2)},\textrm{ when }n=2, \label{g6}\\
h(\kappa):=\int_0^\pi{1-\kappa^2\over 4\pi(1+\kappa^2-2\kappa\cos(\theta))^{3\over 2}}d\theta,\textrm{ when }n\ge 3. \label{g7}
\end{eqnarray}

\begin{theorem}
  \label{thm:HG} 
In the HG phase function in dimension $n=2,3$, the parameter $g(x)$ is uniquely
  determined by the data provided $\sigma_s(x)>0$ for a.e. $x\in X$.
\end{theorem}

Theorem \ref{thm:HG} follows from Theorems \ref{thm:stabsym}, \ref{thm:stabscat}, and from \eqref{g1}, \eqref{g2}, \eqref{g5} and the following Lemma.

\begin{lemma}
\label{hgn=3}
The function $h$ is strictly increasing on $[0,1)$, $\dot h(0)=0$, $\ddot h(0)>0$, $\lim_{g\to 1^-}(1-g)h(g)=c(n)$, where $\dot h(g)={d h\over dg}(g)$ and $c(2)=\pi^{-1}$ and $c(3)=(2\pi)^{-1}$.
\end{lemma}
Considering \eqref{g6}, Lemma \ref{hgn=3} in dimension $n=2$ is trivial. The proof of Lemma \ref{hgn=3} in dimension $n=3$ is given in section \ref{sec:single}.

Moreover we have the following stability estimates. 
\begin{theorem}
\label{thm:stabHG} 
In dimension $n\ge 2$, we have
  \begin{equation}
    \label{eq:HGdim2}
    \begin{array}{ll}
     \|\sigma_g(x) -\tilde\sigma_g(x)\|_{L^1(X)}
   \leq  C\Big(\Big\| \dfrac{(\Gamma_1-\tilde\Gamma_1)(x,x',v')}
   {|\nu(x')\cdot v'| w_n(x,x',v')}
    \Big\|_{\infty}+ \|A-\tilde A\|_{{\mathcal L}(L^1(\Gamma_-,d\xi);L^1(X))}\Big),
    \end{array}
  \end{equation}
  where  
  $w_n(x,x',v')$ is defined  in Theorem \ref{thm:stabscat}, and where the constant $C$ depends on $\delta_0$, $\sigma_0$ and $M$. 

In addition, for the HG phase function in dimension $n=2,3$, we have the following stability estimate: assume $\min(\sigma_s(x),\tilde \sigma_s(x))\ge\sigma_{s,0}>0$ and $(h(g(x)),h(\tilde g(x)))\in W^{1,\infty}(X)^2$, then 
\begin{equation}
    \label{eq:HGdim23}
    \begin{array}{ll}
     \hspace{-.5cm}\|h(g(x))-h(\tilde g(x))\|_{L^1(X)}
   \leq  C\Big(\Big\| \dfrac{(\Gamma_1-\tilde\Gamma_1)(x,x',v')}
   {|\nu(x')\cdot v'| w_n(x,x',v')}
    \Big\|_{\infty}+ \|A-\tilde A\|_
      {{\mathcal L}(L^1(\Gamma_-,d\xi);L^1(X))}\Big),
    \end{array}
\end{equation}
 where the constant $C$ depends on $\delta_0$, $\sigma_0$, $\sigma_{s,0}$, $M$ and $\min(\|h(g)\|_{W^{1,\infty}(X)},\|h(\tilde g)\|_{W^{1,\infty}(X)})$. 
\end{theorem}
Theorem \ref{thm:stabHG} is proved in section \ref{sec:single}.

The sensitivity of the reconstruction of $g(x)$ degrades as $g$
converges to $0$ in the sense that $C(G_0)\to\infty$ as $G_0\to0$,
$G_0=\max(\|g\|_\infty,\|\tilde g\|_\infty)$, where $C$ is the
constant that appears on the right-hand side of \eqref{eq:HGdim23}
when we replace $h(g(x))-h(\tilde g(x))$ by $g-\tilde g$ on the
left-hand side of \eqref{eq:HGdim23}. On the other hand, $C(g_0)\to0$
when $g_0\to 1$, $g_0=\min(\|g\|_\infty,\|\tilde g\|_\infty)$, so that
reconstructions of $g(x)$ are very accurate for $g(x)$ close to $1$,
i.e., in the case of very anisotropic media.


More precisely, using the properties of the function $h$, one can
replace the left-hand side of \eqref{eq:HGdim23} by $\|g^2-\tilde
g^2\|_{L^1(X)}$ (resp. $\|g-\tilde g\|_{L^1(X)}$, resp. $\|{1\over
  1-g}-{1\over {1-\tilde g}}\|_{L^1(X)}$) provided that
$\max(\|g\|_\infty,\|\tilde g\|_\infty)\le G_0<1$ (resp.
$0<g_0<\min(\|g\|_\infty,\|\tilde g\|_\infty)$ and
$\max(\|g\|_\infty,\|\tilde g\|_\infty)\le G_0<1$, resp.
$\min(\|g\|_\infty,\|\tilde g\|_\infty)\ge g_0>0$) for some constant
$G_0$ (resp. $(g_0,G_0)$, resp. $g_0$) and the constant $C$ on the
right hand side of \eqref{eq:HGdim23} then depends also on $G_0$
(resp. $(g_0,G_0)$, resp. $g_0$).
 
\subsection{Reconstructions in the diffusive regime.}
\label{sec:diff}

When scattering is large so that the mean free path $\frac 1\sigma(x)$
is small and intrinsic attenuation $\sigma_a(x)$ is small, then 
radiation inside the domain $X$ is best modeled by a diffusion equation
\begin{equation}
  \label{eq:diff}
  \begin{array}{ll}
  -\nabla\cdot D(x)\nabla I(x) + \sigma_a(x) I(x) =0 \quad & x\in X \\
  I(x) = \phi(x) \quad & x\in\partial X,
  \end{array}
\end{equation}
where $I(x)=\int_{S^{n-1}} u(x,v) dv$ is the spatial density of
photons and $D(x)$ is the diffusion coefficient. We refer the reader to
e.g.  \cite{B-IP-09,dlen6} for references on the diffusion
approximation. When scattering is e.g. isotropic, i.e., when
$k(x,\theta',\theta)=k(x)$, then we find that
$D(x)=\frac{1}{n\sigma_s(x)}$, where $\sigma_s$ is introduced in
\eqref{eq:sigma} and $n$ is the spatial dimension.

When $D(x)$ is known, then the reconstruction of $\sigma_a(x)$ may be
easily obtained by using only one measurement. Indeed, the measurement
$H(x)=\sigma_a(x)I(x)$ so that $I(x)$ may be obtained by solving
\eqref{eq:diff}. Once $I(x)$ is known, it will be positive in $X$
provided that $\phi(x)$ is non trivial and non-negative. Then
$\sigma_a(x)$ is obtained by dividing $H$ by $I$. When $\phi(x)$ is
bounded from below by a positive constant, then we see that the
reconstruction of $\sigma_a$ is unique and clearly stable.

When $(D(x),\sigma_a(x))$ are both unknown, then multiple (at least
two) measurements are necessary. This problem will be analyzed
elsewhere.

\medskip

This concludes the section on the derivation and the display of the
main results. The mathematical proofs are presented in the following
two sections.

\section{Transport equation and estimates}
\label{sec:forward}
In this section, we prove several results on the decomposition of the
albedo operator (Lemmas \ref{lem:dec}, \ref{lem:norm_scat} and
\ref{lem:dec_noyau_stab})
and prove Theorems \ref{thm:decomp} and \ref{thm:asympalpha1}.  

We first recall the well-posedness of the boundary value problem
\eqref{eq:steadytr} (here $\sigma\in L^\infty(X\times \S^{n-1})$,
$k\in L^\infty(X\times\S^{n-1}\times\S^{n-1})$ and $\sigma_a\ge
\sigma_0>0$) and give a decomposition of the albedo operator.  The
boundary value problem \eqref{eq:steadytr} is equivalent to the
integral equation
\begin{equation}
(I-K)u=J\phi\label{eq:int}
\end{equation}
for $u\in L^1(X\times\S^{n-1})$ and $\phi\in L^1(\Gamma_-,d\xi)$, where  $K$ is the bounded operator in $L^1(X\times\S^{n-1})$ defined by 
\begin{equation}
K u=\int_0^{\tau_-(x,v)}E(x,x-tv)\int_{\S^{n-1}}k(x-tv,v', v)u(x-tv,v')dv'dt,\label{eq:K}
\end{equation}
for a.e. $(x,v)\in X\times\S^{n-1}$ and for $u \in L^1(X\times\S^{n-1})$, and $J$ is the bounded operator from $L^1(\Gamma_-,d\xi)$ to $L^1(X\times \S^{n-1})$
defined by 
\begin{equation}
J\phi(x,v)=e^{-\int_0^{\tau_-(x,v)}\sigma(x-pv,v)dp}\phi(x-\tau_-(x,v)v,v),\label{eq:J}
\end{equation}
for a.e. $(x,v)\in X\times\S^{n-1}$.

Since $\sigma_a=\sigma-\sigma_s\ge 0$, it turns out that $I-K$ is
invertible in $L^1(X\times S^{n-1})$ \cite[Lemma 2.4]{BJ-IPI-08}, so
that the albedo operator $A:L^1(\Gamma_-,d\xi)\to L^1(X)$ is well-defined
by \eqref{eq:A} and so that the solution $u$ of \eqref{eq:steadytr}
with boundary condition $\phi\in L^1(\Gamma_-,d\xi)$ satisfies
\begin{equation}
u=\sum_{m=0}^nK^mJ\phi+K^{n+1}(I-K)^{-1}J\phi.\label{dec_sol}
\end{equation}
It follows that
\begin{eqnarray}
A\phi&=&\int_{\Gamma_-}\left(\sum_{m=0}^n\alpha_m(x,x',v')\right)\phi(x',v')d\mu(x')dv'\nonumber\\
&&+\int_{X\times\S^{n-1}}\gamma_{n+1}(x,y,w)((I-K)^{-1}J\phi)(y,w)dy dw,
\label{dec_noyau}
\end{eqnarray}
where $\gamma_m$, $m\ge 0$, is the distributional kernel of  $\bar{K^m}:L^1(X\times\S^{n-1})\to L^1(X)$ defined by
\begin{equation}
\bar {K^m}u(x)=\int_{\S^{n-1}}\sigma_a(x,v)K^mu(x,v)dv,\label{eq:barK}
\end{equation}
for a.e. $x\in X$ and $u\in L^1(X\times \S^{n-1})$, and where $\alpha_m$, $m\ge 0$, is the distributional kernel of $\bar{K^m}J:L^1(\Gamma_-,d\xi)\to L^1(X)$.

We give the explicit expression of the distributional kernels
$\alpha_m$, $m\ge 2$, and $\gamma_m$, $m\ge 3$ in Lemma \ref{lem:dec}
and study the boundedness of $\alpha_m$ in Lemma \ref{lem:norm_scat}
and of $\gamma_{n+1}$ in Lemma \ref{lem:dec_noyau_stab}. We then prove
Theorems \ref{thm:decomp} and \ref{thm:asympalpha1} and Lemmas
\ref{lem:norm_scat} and \ref{lem:dec_noyau_stab}. The proof of Lemma
\ref{lem:dec} is given in the appendix.  For $w\in \R^n$, $w\not=0$,
we set $\hat w:={w\over |w|}$.
\begin{lemma}
\label{lem:dec}
For $m\ge 2$ and a.e. $(z_0,z_m,v_m)\in X\times \Gamma_-$, we have
\begin{eqnarray}
\alpha_m(z_0,z_m,v_m)&=&|\nu(z_m)\cdot v_m|\int_{X^{m-1}}\int_0^{\tau_+(z_m,v_m)}\left[\sigma_a(z_0,v_0)\right.\label{eq:multiple}\\
&&\times {E(z_0,\ldots,z_{m-1},z_m+t'v_m,z_m)k(z_m+t'v_m,v_m,v_{m-1})
\over |z_m+t'v_m-z_{m-1}|^{n-1}\Pi_{i=0}^{m-2}|z_i-z_{i+1}|^{n+1}}\nonumber\\
&&\left.\times\Pi_{i=1}^{m-1}k(z_i,v_i,v_{i-1})\right]_{v_i=\widehat{z_i-z_{i+1}},\ i=0\ldots m-2,\atop  v_{m-1}=\widehat{z_{m-1}-z_m-t'z_m}}dt'dz_1\ldots dz_{m-1}.\nonumber
\end{eqnarray}
For $m\ge 3$ and a.e. $(z_0,z_m,v_m)\in X\times X\times \S^{n-1}$, we have 
\begin{eqnarray}
&&\hspace{-1cm}\gamma_m(z_0,z_m,v_m)=\int_{X^{m-1}}{E(z_0,\ldots,z_m)
\over\Pi_{i=1}^m|z_i-z_{i-1}|^{n-1}}\label{eq:gamma}\\
&&\hspace{-1cm}\times \left[\sigma_a(z_0,v_0)\Pi_{i=1}^mk(z_i,v_i,v_{i-1})\right]_{v_i=\widehat{z_i-z_{i+1}},\ i=0\ldots m-1}dz_1\ldots dz_{m-1}.\nonumber
\end{eqnarray}
\end{lemma}

\begin{lemma}
\label{lem:norm_scat}
For $n= 2$,
\begin{eqnarray}
&&{\alpha_1(x,x',v')\over |\nu(x')\cdot v'|\ln\left({|x-x'-\tau_+(x',v')v'|-(x-x'-\tau_+(x',v')v')\cdot v'\over |x-x'|-(x-x')\cdot v'}\right)}\in L^\infty(X\times\Gamma_-),
\label{eq:norm_simple}\\
&&{\alpha_2(x,x',v')\over |\nu(x')\cdot v'|}\in L^\infty(X\times \Gamma_-).\label{eq:norm_double}
\end{eqnarray}

For $n\ge 3$, we have
\begin{equation}
{|x-x'-((x-x')\cdot v')v'|^{n-1-m}\alpha_m(x,x',v')\over |\nu(x')\cdot v'|}\in L^\infty(X\times\Gamma_-), \ 1\le m\le n-2,
\label{eq:norm_alpham}
\end{equation}
\begin{eqnarray}
&&\hspace{-1cm}{\alpha_{n-1}(x,x',v')\over |\nu(x')\cdot v'|\ln\left({1\over |x-x'-((x-x')\cdot v')v'|}\right)}\in L^\infty(X\times \Gamma_-),\label{eq:norm_alphan-1}\\
&&\hspace{-1cm}{\alpha_n(x,x',v')\over |\nu(x')\cdot v'|}\in L^\infty(X\times \Gamma_-).\label{eq:norm_alphan}
\end{eqnarray}
\end{lemma}

We do not use \eqref{eq:norm_simple} and \eqref{eq:norm_alpham} for
$m=1$ in order to prove \eqref{eq:Gamma2}. However we will use them in
the proof of the stability estimates given in Theorem
\ref{thm:stabscat}.

\begin{lemma}
\label{lem:dec_noyau_stab} 
For $n\ge 2$, we have
\begin{equation}
\gamma_{n+1}\in L^\infty(X\times X\times\S^{n-1}).\label{eq:norm_gamman+1}
\end{equation}
In addition, we have
\begin{equation}
\bar K^{n+1}(I-K)^{-1}J\phi(x)=\int_{\Gamma_-}\Gamma_{n+1}(x,x',v')\phi(x',v')d\mu(x')dv',
\label{dec_noyau_stab}
\end{equation}
for a.e. $x\in X$ and for $\phi\in L^1(\Gamma_-,d\xi)$, where
\begin{equation}
{\Gamma_{n+1}(x,x',v')\over |\nu(x')\cdot v'|}\in L^\infty(X\times \Gamma_-).\label{eq:Gamma_n}
\end{equation}
\end{lemma}

\paragraph{Proof of Theorem \ref{thm:decomp}.}
The equality \eqref{eq:alpha01} for $\alpha_0$ follows from the
definition of the operator $J$ \eqref{eq:J}.  From \eqref{eq:barK},
\eqref{eq:K} and \eqref{eq:J} it follows that
\begin{eqnarray}
\hspace{-1cm}\bar K J\phi(x)&=&\int_{\S^{n-1}}\sigma_a(x,v)\int_0^{\tau_-(x,v)}\int_{\S^{n-1}}E(x,x-tv,x-tv-\tau_-(x-tv,v')v')\nonumber\\
\hspace{-1cm}&&\times k(x-tv,v', v)
\phi(x-tv-\tau_-(x-tv,v')v',v')dv'dtdv,\label{d100}
\end{eqnarray}
for a.e. $x\in X$ and $\phi\in L^1(\Gamma_-,d\xi)$. Performing the change of variables $z=x-tv$ ($dz= t^{n-1}dt dv$, $t=|z-x|$) on the right-hand side of \eqref{d100}, 
we obtain
\begin{eqnarray}
\bar K J\phi(x)&=&\int_{X\times\S^{n-1}}{\sigma_a(x,\widehat{x-z})\over |x-z|^{n-1}}E(x,z,z-\tau_-(z,v')v')\nonumber\\
&&\times k(z,v', \widehat{x-z})
\phi(z-\tau_-(z,v')v',v')dv'dz,\label{d101}
\end{eqnarray}
for a.e. $x\in X$ and $\phi\in L^1(\Gamma_-,d\xi)$. 
Performing the change of variables $z=x'+tv'$ ($x'\in\pa X$, $t>0$, $dz=|\nu(x')\cdot v'|dtd\mu(x')$) on the right hand side of \eqref{d101}, we obtain
\begin{eqnarray}
\bar K J\phi(x)&=&\int_{\Gamma_-}\int_0^{\tau_+(x',v')}{\sigma_a(x,\widehat{x-x'-t'v'})\over |x-x'-t'v'|^{n-1}}E(x,x'+t'v',x')\nonumber\\
&&\times k(x'+t'v',v', \widehat{x-x'-t'v'})
\phi(x',v')dtd\xi(x',v'),\label{d106}
\end{eqnarray}
for a.e. $x\in X$ and $\phi\in L^1(\Gamma_-,d\xi)$, which yields \eqref{eq:alpha01} for $\alpha_1$.

Now set $\Gamma_2:=\sum_{m=2}^n\alpha_m+\Gamma_{n+1}$ when $n\ge 2$. Taking account of Lemma \ref{lem:norm_scat}  \eqref{eq:norm_double}--\eqref{eq:norm_alphan} and Lemma \ref{lem:dec_noyau_stab} \eqref{eq:Gamma_n}, we obtain \eqref{eq:Gamma2}.\hfill $\Box$

\paragraph{Proof of Theorem \ref{thm:asympalpha1}.}
We assume that
$\sigma \in {\cal C}_b(X\times\S^{n-1})$ and $k\in {\cal C}_b(X\times\S^{n-1}\times\S^{n-1})$.
Let $(x',v')\in \Gamma_-$ and let $t_0'\in (0,\tau_+(x',v'))$ and let ${v'}^\bot$ be such that $v'\cdot {v'}^\bot=0$. Set $x=x'+t_0'v'$. From \eqref{eq:alpha01}, it follows that
\begin{equation}
{\alpha_1(x+\ep{v'}^\bot,x',v')\over|\nu(x')\cdot v'|}=(L_++L_-)(\ep),
  \qquad \mbox{ where }\label{pp1}
\end{equation}
\begin{equation}
L_+(\ep):=\int_0^{t_0'}{E(x+\ep{v'}^\bot,x'+t'v',x')
\over ((t'-t_0')^2+\ep^2)^{n-1\over 2}}\sigma_a(x+\ep {v'}^\bot,v_{t',\ep})k(x'+t'v',v',v_{t',\ep})dt',\label{pp1a}
\end{equation}
\begin{equation}
L_-(\ep):=\int_{t_0'}^{\tau_+(x',v')}{E(x+\ep{v'}^\bot,x'+t'v',x')
\over ((t'-t_0')^2+\ep^2)^{n-1\over 2}}\sigma_a(x+\ep {v'}^\bot,v_{t',\ep})k(x'+t'v',v',v_{t',\ep})dt',\label{pp1b}
\end{equation}
for $\ep\in (0,\tau_+(x,{v'}^\bot))$, where $v_{t',\ep}=\widehat{{(t_0'-t')\over \ep}v'+{v'}^\bot}$ for $t'\in \R$. 

We prove \eqref{eq:asyalpha1} for $n=2$.
Consider the function $\arcsinh:\R\to\R$ defined by
$\arcsinh(y):=\ln(y+\sqrt{1+y^2})$, for $y\in \R$.
Then performing the change of variables $\eta:={\arcsinh({t_0'-t'\over \ep})\over \arcsinh({t_0'\over \ep})}$ on the right hand side of \eqref{pp1a} and performing the change of variables $\eta:={\arcsinh({t'-t_0'\over \ep})\over \arcsinh({\tau_+(x',v')-t_0'\over \ep})}$ on the right hand side of \eqref{pp1b}, we obtain
\begin{equation}
L_\pm(\ep):=\arcsinh({s_\pm\over \ep})L_\pm'(\ep),
\label{pp2a}
\end{equation}
for $\ep\in (0,\tau_+(x,{v'}^\bot))$, where
\begin{eqnarray}
\hspace{-2cm}&&s_+:=t_0',\ s_-:=\tau_+(x',v')-t_0',\label{pp2aaa}\\
\hspace{-2cm}L_\pm'(\ep)&:=&\int_0^1E(x+\ep{v'}^\bot,x'+t'_\pm(\eta,\ep)v',x')\nonumber\\&& 
\quad \sigma_a(x+\ep {v'}^\bot,v_\pm(\eta,\ep))k(x'+t'_\pm(\eta,\ep)v',v',v_\pm(\eta,\ep))d\eta\label{pp2aa}\\
\hspace{-2cm}t'_\pm(\eta,\ep)&:=&t_0'\mp\ep\sinh\left(\eta\arcsinh({s_\pm\over \ep})\right),\label{pp2b}\\
\hspace{-2cm}v_\pm(\eta,\ep)&:=&{\pm\sinh\left(\eta\arcsinh({s_\pm\over \ep})\right)v'+{v'}^\bot\over \sqrt{\sinh\left(\eta\arcsinh({s_\pm\over \ep})\right)^2+1}},\label{pp2c}
\end{eqnarray}
for $\eta\in (0,1)$ (we recall that $\sinh(y)={e^y-e^{-y}\over 2}$, $y\in \R$).
Note that using the definition of $\sinh$ and $\arcsinh$, we obtain
\begin{equation}
|\ep\sinh\left(\eta\arcsinh({s_\pm\over \ep})\right)|\le {e^{\eta\ln(s_\pm+\sqrt{s_\pm^2+1})}\ep^{1-\eta}\over 2},\label{pp4}
\end{equation}
for $\eta\in (0,1)$.
Therefore using \eqref{pp2b} we obtain
\begin{equation}
t_\pm'(\eta,\ep)\to t_0',\textrm{ as }\ep\to 0^+\label{pp6}
\end{equation}
for $\eta\in (0,1)$ and $i=1,2$.
Note also that from \eqref{pp2c} it follows that
\begin{equation}
v_\pm(\eta,\ep)={\pm v'+\sinh\left(\eta\arcsinh({s_\pm\over \ep})\right)^{-1}{v'}^\bot\over \sqrt{1+\sinh\left(\eta\arcsinh({s_\pm\over \ep})\right)^{-2}}}\underset{\ep \to 0^+}
{\longrightarrow} \pm v',\label{pp5a}\\
\end{equation}
for $\eta\in (0,1)$ (we used the limit $\sinh\left(\eta\arcsinh({s\over \ep})\right)\to +\infty$ as $\ep\to 0^+$ which holds for any positive real numbers $s$ and $\eta$).

Using \eqref{pp2aa}, \eqref{pp6}, \eqref{pp5a} and continuity and boundedness of $\sigma$ and $k$ and $\sigma_a$, and using Lebesgue dominated convergence theorem, we obtain
\begin{equation} 
L_\pm'(\ep)\to\sigma_a(x,\pm v')E(x,x')k(x'+t_0'v',v',\pm v')\textrm{ as }\ep\to 0^+.\label{pp7a}
\end{equation}
Finally note that
\begin{equation}
\arcsinh({s_\pm\over \ep})=\ln\big({1\over \ep}\big)+o\Big({1\over \ep}\Big), \textrm{ as }\ep\to 0^+. \label{pp8}
\end{equation}
Combining \eqref{pp1}, \eqref{pp2a},  \eqref{pp7a}  and \eqref{pp8}, we obtain \eqref{eq:asyalpha1} for $n=2$.

We prove \eqref{eq:asyalpha1} for $n\ge 3$. 
Performing the change of variables $\eta={t'-t_0'\over \ep}$ on the
right-hand side of \eqref{pp1a}, we obtain
\begin{equation}
\begin{array}{ll}
L_+(\ep)=\ep^{2-n}\dint_{-{t_0'\over \ep}}^0E(x+\ep{v'}^\bot,x'+t'_3(\ep,\eta)v',x')
\\
{\sigma_a(x+\ep {v'}^\bot,v)
\over \sqrt{\eta^2+1}^{n-1}}k(x'+t'_3(\ep,\eta)v',v',v)_{|v=\widehat{{-\eta v'+{v'}^\bot}}}d\eta,\label{pp9a}
\end{array}
\end{equation}
for $\ep\in (0,\tau_+(x,{v'}^\bot))$, where
$t'_3(\ep,\eta)=t_0'+\ep\eta$ for $\eta\in \R$. Note that
\begin{equation}
t'_3(\ep,\eta)\to t_0',\textrm{ as }\ep\to 0^+.\label{pp9b}
\end{equation}
Therefore, using the Lebesgue dominated convergence theorem and continuity
and boundedness of $(\sigma, k, \sigma_a)$, we obtain
\begin{equation}
L_+(\ep)=\ep^{2-n}E(x,x')\int_{-\infty}^0{\sigma_a(x,v)k(x,v',v)_{|v=\widehat{{-\eta v'+{v'}^\bot}}}
\over \sqrt{\eta^2+1}^{n-1}}d\eta+o(\ep^{2-n}),\label{10}
\end{equation}
as $\ep\to 0^+$.  Similarly performing the change of variables
$\eta={t'-t_0'\over \ep}$ on the right hand side of \eqref{pp1b}, and
using Lebesgue dominated convergence theorem and continuity and
boundedness of $(\sigma, k, \sigma_a)$, we obtain
\begin{equation}
L_-(\ep)=\ep^{2-n}E(x,x')\int_0^{+\infty}{\sigma_a(x,v)k(x,v',v)_{|v=\widehat{{-\eta v'+{v'}^\bot}}}
\over \sqrt{\eta^2+1}^{n-1}}d\eta+o(\ep^{2-n}),\label{pp11b}
\end{equation}
as $\ep\to 0^+$.  Note that performing the change of variables
$\cos(\theta)={-\eta\over \sqrt{1+\eta^2}}$, $\theta\in (0,\pi)$,
($\eta=-{\cos(\theta)\over \sin(\theta)}$, $d\eta={1\over
  \sin(\theta)^2}d\theta$) we have
\begin{equation}
\int_{-\infty}^{+\infty}{\sigma_a(x,v)k(x,v',v)_
   {|v=\widehat{{-\eta v'+{v'}^\bot}}}
\over \sqrt{\eta^2+1}^{n-1}}d\eta=\int_0^\pi\sin(\theta)^{n-3}
\sigma_a(x,v(\theta))k(x,v',v(\theta))d\theta.\label{pp12}
\end{equation}
Adding \eqref{10} and \eqref{pp11b} and using \eqref{pp12} and \eqref{pp1}, we obtain \eqref{eq:asyalpha1} for $n\ge 3$.\hfill$\Box$

\paragraph{Proof of Lemma  \ref{lem:norm_scat}.}
  First note that from \eqref{eq:alpha01} and
  \eqref{eq:multiple}, it follows that
\begin{equation}
\alpha_m(x,x',v')\le \|\sigma_a\|_\infty\|k\|_{\infty}^m|\nu(x')\cdot v'|I_{m,n}(x,x',v'),\label{p1}
\end{equation}
for a.e. $(x,x',v')\in X\times \Gamma_-$ and for $m\in \N$, $m\ge 1$, where
\begin{eqnarray}
\hspace{-1cm}I_{1,n}(z_0,x',v')&=&\int_0^{\tau_+(x',v')}{dt'\over |z_0-x'-t'v'|^{n-1}},\label{p2a}\\
\hspace{-1cm}I_{m+1,n}(z_0,x',v')&=&\int_{X^m}\int_0^{\tau_+(x',v')}
  \hspace{-.5cm}{dt'dz_1\ldots dz_m\over |x'+t'v'-z_m|^{n-1}\Pi_{i=1}^m|z_i-z_{i-1}|^{n-1}},\label{p2b}
\end{eqnarray}
for $(z_0,x',v')\in X\times \Gamma_-$ and $m\ge 1$.

We prove \eqref{eq:norm_simple} and \eqref{eq:norm_double}.
Let $n=2$.
Let $(x,x',v')\in X\times \Gamma_-$ be such that $x\not=x'+\lambda v'$ for any $\lambda\in \R$. Set $(w)_\bot:=w-(w\cdot v')v'$ for any $w\in \R^n$.
Using \eqref{p2a} and using the equality $|x-x'-t'v'|^2=(t'-(x-x')\cdot v')^2+|(x-x')_\bot|^2$, we obtain
\begin{eqnarray}
\hspace{-1cm}I_{1,n}(x,x',v')&=&\int_{-{(x-x')\cdot v'}}^{\tau_+(x',v')-(x-x')\cdot v'}{1\over (|(x-x')_\bot|^2+t^2)^{n-1\over 2}}dt,\label{p3}\\
\hspace{-1cm}&=&\ln\Big({|x-x'-\tau_+(x',v')v'|- (x-x'-\tau_+(x',v')v')\cdot v'\over |x-x'|-(x-x')\cdot v'}\Big),\label{p4}
\end{eqnarray}
where we used that $\int_0^x{dt\over
  \sqrt{1+t^2}}=\ln\left(x+\sqrt{1+x^2}\right)$.
Estimate
\eqref{eq:norm_simple} follows from \eqref{p1} and \eqref{p4}.

Let $(x,x',v')\in X\times\Gamma_-$ be such that $x\not= x'+\lambda v'$ for any $\lambda\in \R$. 
Using \eqref{p4} (with ``$x=z$'') and \eqref{p2b}, we obtain
\begin{eqnarray} 
I_{2,n}(x,x',v')&=&\int_X{\ln\left({|z-x'-\tau_+(x',v')v'|- (z-x'-\tau_+(x',v')v')\cdot v'|\over |z-x'|-(z-x')\cdot v'}\right)\over |x-z|}dz\nonumber\\
&=&I_{2,n}'(x,x'+\tau_+(x',v')v',v')-I_{2,n}'(x,x',v'),\label{p7}
\end{eqnarray}
where 
\begin{equation}
I_{2,n}'(x,a,v')=\int_X{\ln(|z-a|- (z-a)\cdot v')\over |x-z|}dz,\label{p8}
\end{equation}
for $a\in \bar X$.
We prove
\begin{equation}
\sup_{(x,a,v')\in X\times\bar X\times\S^{n-1}}|I_{2,n}'(x,a,v')|<\infty.\label{p9}
\end{equation}
Then estimate \eqref{eq:norm_double} follows from \eqref{p1}, \eqref{p7} and \eqref{p9}.

Let $(x,a,v')\in X\times\bar X\times\S^{n-1}$. Note that by using
\eqref{p8}, we obtain
\begin{eqnarray}
|I_{2,n}'(x,a,v')|&\le& C\int_X{(|z-a|+(z-a)\cdot v')^{1\over 8}\over |x-z|(|z-a|^2-((z-a)\cdot v')^2)^{1\over 8}}dz\nonumber\\
&\le&(2D)^{1\over 8}C\int_X{1\over |x-z|(|z-a|^2-((z-a)\cdot v')^2)^{1\over 8}}dz,\label{p10}
\end{eqnarray}
where $C:=\sup_{r\in (0,6D)}r^{1\over 8}|\ln(r)|<\infty$ and $D$ denotes the diameter of $X$.
Consider ${v'}^\bot\in \S^{n-1}$ a unit vector orthogonal to $v'$ and perform the change of variable $z=a+\lambda_1 v'+\lambda_2 {v'}^\bot$, then we obtain
\begin{equation}
|I_{2,n}'(x,a,v')|\le (2D)^{1\over 8}C\int_{[-D,D]^2}{d\lambda_1d\lambda_2\over ((\lambda_{1,x}-\lambda_1)^2+(\lambda_{2,x}-\lambda_2)^2)^{1\over 2}|\lambda_2|^{1\over 4}},\label{p11}
\end{equation}
where $\lambda_{1,x}v'+\lambda_{2,x}{v'}^\bot=x-a$.
Then note that 
\begin{equation}
\int_{-D}^D{d\lambda_1\over ((\lambda_{1,x}-\lambda_1)^2+(\lambda_{2,x}-\lambda_2))^{1\over2}}
= \ln\left({D-\lambda_{1,x}+((\lambda_2-\lambda_{2,x})^2+(D-\lambda_{1,x})^2)^{1\over 2}
\over -D-\lambda_{1,x}+((\lambda_2-\lambda_{2,x})^2+(D+\lambda_{1,x})^2))^{1\over 2}}\right),\label{p12}
\end{equation}
Combining \eqref{p11} and \eqref{p12}, we obtain
\begin{eqnarray}
&&\hspace{-1cm}|I_{2,n}'(x,a,v')|\le C'\int_{-D}^D{(D-\lambda_{1,x}+((\lambda_2-\lambda_{2,x})^2+(D-\lambda_{1,x})^2)^{1\over 2})^{1\over 8}d\lambda_2\over |\lambda_2|^{1\over 4}
 (-(D+\lambda_{1,x})+\sqrt{(\lambda_2-\lambda_{2,x})^2+(D+\lambda_{1,x})^2})^{1\over 8}}\nonumber\\
&&\hspace{-1cm}\le C''\int_{-D}^D{(D+\lambda_{1,x}+((\lambda_2-\lambda_{2,x})^2+(D+\lambda_{1,x})^2)^{1\over 2})^{1\over 8}\over |\lambda_2|^{1\over 4}
 |\lambda_2-\lambda_{2,x}|^{1\over 4}}d\lambda_2\nonumber\\
&&\hspace{-1cm}\le C'''\int_{-D}^D{d\lambda_2\over |\lambda_2|^{1\over 4}|\lambda_2-\lambda_{2,x}|^{1\over 4}}
=C'''|\lambda_{2,x}|^{1\over 2}\int_{-{D\over |\lambda_{2,x}|}}^{D\over |\lambda_{2,x}|}{ds\over |s|^{1\over 4}|s-1|^{1\over 4}},\label{p13}
\end{eqnarray}
where $C':=(2D)^{1\over 8}C^2$, $C'':=(12D^2)^{1\over 8}C^2$ and  $C''':=(72D^3)^{1\over 8}C^2$. 
Finally note that
\begin{equation}
|\lambda_{2,x}|^{1\over 2}\int_{|s|\le{D\over |\lambda_{2,x}|}}{1\over |s|^{1\over 4}|s-1|^{1\over 4}}ds\le C_1(\lambda_{2,x})+C_2(\lambda_{2,x}),
\quad \mbox{ where } \label{p14}
\end{equation}
\begin{eqnarray}
\hspace{-1cm}C_1(\lambda_{2,x})&:=&|\lambda_{2,x}|^{1\over 2}
 \int_{|s|\le 2}{|s|^{-{1\over 4}}|s-1|^{-{1\over 4}}}ds\le 
  D^{1\over 2}\int_{|s|\le 2}{|s|^{-{1\over 4}}|s-1|^{-{1\over 4}}}ds
  \,\label{p15a}\\
\hspace{-1cm}C_2(\lambda_{2,x})&:=&|\lambda_{2,x}|^{1\over 2}
 \int_{2\le|s|\le\max(2,{D\over |\lambda_{2,x}|})}
  \hspace{-1.5cm} {|s|^{-{1\over 4}}|s-1|^{-{1\over 4}}}ds
\le |\lambda_{2,x}|^{1\over 2}2^{1\over 4}\int_{2\le|s|\le\max(2,{D\over |\lambda_{2,x}|})}
   \hspace{-1cm}{|s|^{-{1\over 2}}}ds\nonumber\\
\hspace{-1cm}&\le & 2^{9\over 4}\Big(|\lambda_{2,x}|\max(2,{D\over |\lambda_{2,x}|})\Big)^{1\over 2}\le 2^{11\over 4}D^{1\over 2}.\label{p15b}
\end{eqnarray}
Combining \eqref{p13}--\eqref{p15b} we obtain \eqref{p9}.

The statements \eqref{eq:norm_alpham}--\eqref{eq:norm_alphan} follow from \eqref{p1} and the following statements \eqref{eq:norm_Im}--\eqref{eq:norm_In}
\begin{eqnarray}
&&\hspace{-1cm}|x-x'-((x-x')\cdot v')v'|^{n-1-m}I_{m,n}(x,x',v')\in L^\infty(X\times\Gamma_-),
\label{eq:norm_Im}\\
&&\hspace{-1cm}{I_{n-1,n}(x,x',v')\over \ln\left({1\over |x-x'-((x-x')\cdot v')v'|}\right)}\in L^\infty(X\times \Gamma_-),\label{eq:norm_In-1}\\
&&\hspace{-1cm}I_{n,n}(x,x',v')\in L^\infty(X\times \Gamma_-),\label{eq:norm_In}
\end{eqnarray} 
for $(m,n)\in \N\times \N$, $n\ge 3$, $1\le m\le n-2$ and where $I_{m,n}$ is defined by \eqref{p2a} and \eqref{p2b}.

We prove \eqref{eq:norm_Im}--\eqref{eq:norm_In}, which will complete the proof of Lemma \ref{lem:norm_scat}.
We proceed by induction on $m$.
We prove \eqref{eq:norm_Im} for $m=1$. 
Let $n\ge 3$. Note that formula \eqref{p3} still holds. 
Note also that 
\begin{equation}
|w-\lambda v'|^2=|w_\bot|^2+|w\cdot v'-\lambda|^2\ge 2^{-1}(|w_\bot|+|w\cdot v'-\lambda|)^2,\label{p18}
\end{equation}
for $(w,\lambda)\in \R^n\times (-D,D)$,\ $|w|\le 2D$. Therefore
\begin{equation}
\int_{-2D}^{2D}{d\lambda\over |w-\lambda v'|^{n-1}}\le \int_{-2D}^{2D}{2^{n-1\over 2}d\lambda\over (|w_\bot|+|\lambda|)^{n-1}}\le {2^{n+1\over 2}\over (n-2)|w_\bot|^{n-2}}.\label{p19}
\end{equation}
Thus \eqref{eq:norm_Im} for $m=1$ follows from \eqref{p19} and \eqref{p3}.

Let $m\in \N$, $m\ge 1$ be such that \eqref{eq:norm_Im}--\eqref{eq:norm_In} hold for any $n\ge 3$.
We prove that \eqref{eq:norm_Im}--\eqref{eq:norm_In} hold for any $n\ge 3$ and for ``$m$''$=m+1$.
Let $(x,x',v')$ be such that $x\not =x'+\lambda v'$ for any $\lambda \in \R$. From \eqref{p2a} and \eqref{p2b} it follows that
\begin{equation}
I_{m+1,n}(x,x',v')\le \int_X{I_{m,n}(z,x',v')\over |x-z|^{n-1}}dz.\label{p16}
\end{equation}
Assume that $m+1\le n-1$. Then from \eqref{p16} and \eqref{eq:norm_Im} for $(m,n)$ it follows that there exists a constant $C$ (which does not depend on $(x,x',v')$) such that
\begin{equation}
I_{m+1,n}(x,x',v')\le C\int_X{1\over |x-z|^{n-1}|(z-x')_\bot|^{n-1-m}}dz.\label{p16b}
\end{equation}
Performing the change of variables $z-x'=z'+\lambda v'$, $z'\cdot
v'=0$, we obtain
\begin{equation}
I_{m+1,n}(x,x',v')\le C\int_{z'\cdot v'=0\atop |z'|\le D}\left(\int_{-D}^D{d\lambda\over |x-x'-z'-\lambda v'|^{n-1}}\right){dz'\over |z'|^{n-1-m}}.\label{p17}
\end{equation}
Combining \eqref{p17} and \eqref{p19} (with ``$w=x-x'-z'$''), we obtain
\begin{equation}
I_{m+1,n}(x,x',v')\le {2^{n+1\over 2}C\over n-2}I_{m+1,n}''(x,x',v'),
\quad \mbox{ where }\label{p20}
\end{equation}
\begin{eqnarray}
I_{m+1,n}''(x,x',v')&:=&\int_{z'\cdot v'=0\atop |z'|\le D}{dz'\over |(x-x')_\bot-z'|^{n-2}|z'|^{n-1-m}}\nonumber\\
&=&\int_{z'\in B_{n-1}(0,D)}{dz'\over ||(x-x')_\bot|e_1-z'|^{n-2}|z'|^{n-1-m}},\label{p20b}
\end{eqnarray}
and $e_1=(1,0,\ldots,0)\in \R^{n-1}$ and $B_{n-1}(0,D)$ denotes the
Euclidean ball of $\R^{n-1}$ of center $0$ and radius $D$. Using
spherical coordinates $z'=|(x-x')_\bot|e_1+r\Omega$, $(r,\Omega)\in
(0,+\infty)\times\S^{n-2}$ and
$\Omega=(\sin(\theta),\cos(\theta)\Theta)$ ($(\theta,\Theta)\in
(-{\pi\over 2},{\pi\over 2}) \times \S^{n-3}$) we obtain
\begin{eqnarray}
&&\hspace{-1.5cm}I_{m+1,n}''(x,x',v')\le |\S^{n-3}|\int_{-{\pi\over 2}}^{\pi\over 2}\cos^{n-3}(\theta)\nonumber\\
&&\hspace{-1.5cm}\times\Big(\int_0^{2D}{dr\over (|r+|(x-x')_\bot|\sin(\theta)|^2
+|(x-x')_\bot|^2\cos(\theta)^2)^{n-1-m\over 2}}\Big) d\theta\label{p21}
\end{eqnarray}
(by convention $|\S^0|:=2$).  Note that by performing the change of
variables ``$r=r+|(x-x')_\bot|\sin (\theta)$'' and using the estimate
$a^2+b^2\ge 2^{-1}(a+b)^2$, we obtain
$$
\int_0^{2D}{dr\over (|r+|(x-x')_\bot|\sin(\theta)|^2
+|(x-x')_\bot|^2\cos(\theta)^2)^{n-1-m\over 2}}
$$
\begin{eqnarray}
\hspace{-1cm}&&\le\int_{-3D}^{3D}{2^{n-1\over 2}dr\over (|r|+|(x-x')_\bot|\cos(\theta))^{n-1-m}}=\int_0^{3D}{2^{n+1\over 2}dr\over (r+|(x-x')_\bot|\cos(\theta))^{n-1-m}}\nonumber\\
\hspace{-1cm}&&\le\left\lbrace
\begin{array}{l} 
\displaystyle {2^{n+1\over 2}\over (n-m-2)(|(x-x')_\bot|\cos(\theta))^{n-2-m}},\textrm{ if }m+1<n-1,\\
\displaystyle 2^{n+1\over 2}\ln\left({3D+|(x-x')_\bot|\cos(\theta)\over |(x-x')_\bot|\cos(\theta)}\right),\textrm{ if }m+1=n-1, 
\end{array}
\right.\label{p22}
\end{eqnarray}
for $\theta\in (-{\pi\over 2},{\pi\over 2})$.
Assume $m+1<n-1$. Then combining \eqref{p21} and \eqref{p22}, we obtain
\begin{equation}
I_{m+1,n}''(x,x',v')\le{2^{n+1\over 2}|\S^{n-3}|\over (n-2-m)|(x-x')_\bot|^{n-2-m}}\int_{-{\pi\over 2}}^{\pi\over 2}\cos^{m-1}(\theta)d\theta.\label{p23}
\end{equation}
Therefore using also \eqref{p20} we obtain that \eqref{eq:norm_Im} holds for ``$m$''$=m+1<n-1$. 
Assume $m+1=n-1$. Then note that
\begin{eqnarray}
&&\ln\left({3D+|(x-x')_\bot|\cos(\theta)\over |(x-x')_\bot|\cos(\theta)}\right)\le\ln\left({4D\over |(x-x')_\bot|\cos(\theta)}\right)\nonumber\\
&&\le\ln(4D)+\ln({1\over |(x-x')_\bot|})-\ln(\cos(\theta)),\label{p24}
\end{eqnarray}
for $\theta\in (-{\pi\over 2},{\pi\over 2})$.
Combining \eqref{p21}, \eqref{p22} and \eqref{p24}, we obtain 
\begin{equation}
I_{m+1,n}''(x,x',v')\le 2^{n+1\over 2} |\S^{n-3}|\left(C_1+C_2\ln({1\over |(x-x')_\bot|})\right),\label{p25}
\end{equation}
where $C_1:=\int_{-{\pi\over 2}}^{\pi\over 2}\cos^{n-3}(\theta)(\ln(4D)-\ln(\cos(\theta)))d\theta<\infty$ and $C_2:=\int_{-{\pi\over 2}}^{\pi\over 2}\cos^{n-3}(\theta)d\theta$.
Therefore using also \eqref{p20}, we obtain that \eqref{eq:norm_In-1} holds for ``$m$''$=m+1=n-1$. 

Assume that $m+1=n$. From \eqref{p16} and \eqref{eq:norm_In-1} for $(n-1,n)$ it follows that 
there exists a constant $C$ (which does not depend on $(x,x',v')$) such that
\begin{equation}
I_{m+1,n}(x,x',v')\le C\int_X{\left|\ln\left({1\over |(z-x')_\bot|}\right)\right|\over |x-z|^{n-1}}dz.\label{p26}
\end{equation}
Performing the change of variables $z-x'=z'+\lambda v'$, $z'\cdot v'=0$, we obtain
\begin{equation}
I_{m+1,n}(x,x',v')\le C\int_{z'\cdot v'=0\atop |z'|\le D}\left(\int_{-D}^D{d\lambda\over |x-x'-z'-\lambda v'|^{n-1}}\right)
\left|\ln\left({1\over |z'|}\right)\right|dz'.\label{p27}
\end{equation}
Combining \eqref{p27} and \eqref{p19} (with ``$w=x-x'-z'$''), we obtain
\begin{equation}
I_{m+1,n}(x,x',v')\le {2^{n+1\over 2}C\over n-2}I_{n,n}''(x,x',v'),\label{p30}
\end{equation}
where
\begin{eqnarray}
\hspace{-2.5cm}
I_{n,n}''(x,x',v'):=\int_{z'\cdot v'=0\atop |z'|\le D}{\left|\ln\left({1\over |z'|}\right)\right|\over |(x-x')_\bot-z'|^{n-2}}dz'\label{p30b}\\
\hspace{-2.5cm}
=\int_{z'\in B_{n-1}(0,D)}{\left|\ln\left({1\over |z'|}\right)\right|\over ||(x-x')_\bot|e_1-z'|^{n-2}}dz'
\le\int_{z'\in B_{n-1}(0,D)}{C'\over |z'|^{1\over 2}||(x-x')_\bot|e_1-z'|^{n-2}}dz',\nonumber
\end{eqnarray}
and $C':=\sup_{r\in (0,D)}r^{1\over 2}|\ln(r)|$,
$e_1=(1,0,\ldots,0)\in \R^{n-1}$, and where $B_{n-1}(0,D)$ denotes the
Euclidean ball of $\R^{n-1}$ of center $0$ and radius $D$.  Using
spherical coordinates $z'=|(x-x')_\bot|e_1+r\Omega$, $(r,\Omega)\in
(0,+\infty)\times\S^{n-2}$ and
$\Omega=(\sin(\theta),\cos(\theta)\Theta)$ ($(\theta,\Theta)\in
(-{\pi\over 2},{\pi\over 2}) \times \S^{n-3}$), and using the estimate
$||(x-x')_\bot|e_1+r\Omega|\ge |r+|(x-x')_\bot|\sin(\theta)|$, we
obtain
\begin{equation}
I_{n,n}''(x,x',v')\le |\S^{n-3}|\int_{-{\pi\over 2}}^{\pi\over 2}\cos^{n-3}(\theta)\left(\int_0^{2D}{C'\over |r+|(x-x')_\bot|\sin(\theta)|^{1\over 2}}dr
\right) d\theta.\label{p31}
\end{equation}
Note that
\begin{eqnarray}
\int_0^{2D}{C'\over |r+|(x-x')_\bot|\sin(\theta)|^{1\over 2}}dr\le \int_{-3D}^{3D}{C'\over |r|^{1\over 2}}dr<\infty,\label{p32}
\end{eqnarray}
for $\theta\in (-{\pi\over 2},{\pi\over 2})$.
Combining \eqref{p30}, \eqref{p31} and \eqref{p32}, we obtain
\begin{equation}
I_{m+1,n}(x,x',v')\le {2^{n+1\over 2}CC'|\S^{n-3}|\over n-2}\int_{-{\pi\over 2}}^{\pi\over 2}\cos^{n-3}(\theta)d\theta \int_{-3D}^{3D}r^{-{1\over 2}}dr.\label{p33}
\end{equation}
Therefore \eqref{eq:norm_In} holds for ``$m$''$=m+1=n$. $\square$

\paragraph{Proof of Lemma \ref{lem:dec_noyau_stab}. }
We first prove the estimates \eqref{s3b} and \eqref{s3} given below
\begin{eqnarray}
\int_X{dz_1\over |z_0-z_1||z_1-z|^{n-1}}&\le& C_2-C_2'\ln(|z_0-z|),\textrm{ when }n=2\label{s3b}\\
\int_X{dz_1\over |z_0-z_1|^m|z_1-z|^{n-1}}&\le& {C_n\over |z_0-z|^{m-1}}, \textrm{ when }n\ge 3,\label{s3}
\end{eqnarray}
for $(z_0,z)\in X^2$ and for  $m\in \N$ such that $z_0\not=z$ and $2\le m\le n-1$, where the positive constants $C_n$, $C_2'$ do not depend on $(z_0,z)$. 

Let $(z_0,z)\in X^2$ and let $m\in \N$ be such that $z_0\not=z$ and
$1\le m\le n-1$. Performing the change of variables
$z_1=z+r_1\Omega_1$, $(r_1,\Omega_1)\in (0,D)\times \S^{n-1}$ (where
$D$ denotes the diameter of $X$), we obtain
\begin{eqnarray}
\hspace{-2.5cm}\int_X{dz_1\over |z_0-z_1|^m|z_1-z|^{n-1}}&\le&\!\!\int_0^{D}
\!\!\int_{\S^{n-1}}\!{d\Omega_1\over |r\Omega-r_1\Omega_1|^m}dr_1
=\!\!\int_0^{D}\!\!
\int_{\S^{n-1}}\!{d\Omega_1\over |re_1-r_1\Omega_1|^m}dr_1\label{s4}
\end{eqnarray}
where $e_1=(1,0,\ldots,0)\in \R^n$ (we also used a rotation that maps $\Omega$ to the vector $e_1$) and
\begin{equation} 
r=|z-z_0|,\ \Omega={z_0-z\over |z_0-z|}.\label{s4b}
\end{equation}
Performing the change of variables
$\Omega_1=(\sin(\theta_1),\cos(\theta_1)\Theta_1)$,
$(\theta_1,\Theta_1)\in (-{\pi\over 2},{\pi\over 2})\times \S^{n-2}$),
on the right hand side of \eqref{s4} we obtain
\begin{equation}
\int_0^{D}\int_{\S^{n-1}}{d\Omega_1\over |re_1-r_1\Omega_1|^m}dr_1=c(n)\int_0^{D}\int_{-{\pi \over 2}}^{\pi \over 2}{\cos(\theta_1)^{n-2}d\theta_1\over 
(r^2+r_1^2-2rr_1\sin(\theta_1))^{m\over 2}}dr_1,\label{s5}
\end{equation}
where $c(n):=|\S^{n-2}|$ (by convention $c(2):=2$). 

Consider the case $n=2$ and $m=1$.  Using \eqref{s5} and the estimate
$(r2+r_12-2rr_1\sin(\theta_1))^{1\over 2}\ge 2^{-{1\over
    2}}(|r_1-r\sin(\theta_1)|+r|\cos(\theta_1)|)$, we obtain
\begin{equation}
S(r):=\int_0^{D}\int_{-{\pi \over 2}}^{\pi \over 2}{d\theta_1\over
(r^2+r_1^2-2rr_1\sin(\theta_1))^{m\over 2}}dr_1\label{s77}
\end{equation}
$$
\le\int_0^{2\pi}\int_0^{D}{\sqrt{2}dr_1\over |r_1-r\sin(\theta_1)|+r|\cos(\theta_1)|}d\theta_1
$$
$$
\le\int_0^{2\pi}\int_{-2D}^{2D}{\sqrt{2}dr_1\over |r_1|+r|\cos(\theta_1)|}d\theta_1
\le\int_0^{2\pi}\int_0^{2D}{2^{3\over 2}dr_1\over r_1+r|\cos(\theta_1)|}d\theta_1
$$
\begin{equation}
\le 2^{3\over 2}\int_0^{2\pi}\ln\left({2D+r|\cos(\theta_1)|\over r|\cos(\theta_1)|}\right)d\theta_1
\le2^{3\over 2}\int_0^{2\pi}\ln\left({4D\over r|\cos(\theta_1)|}\right)d\theta_1\le C_1-2^{7\over 2}\pi\ln(r),
\label{s7.b}
\end{equation}
where $C_1:=2^{3\over
  2}\int_0^{2\pi}(\ln(4D)-\ln(|\cos(\theta_2)|))d\theta_2<\infty$.
Estimate \eqref{s3b} follows from \eqref{s4}, \eqref{s5} and
\eqref{s7.b}.


Consider the case $n\ge 3$ and $2\le m\le n-1$.
Note that 
\begin{equation}
r\cos(\theta_1)\le \sqrt{r^2+r_1^2-2rr_1\sin(\theta_1)}, \label{s6}
\end{equation}
for $(r,r_1,\theta_1)\in (0,+\infty)^2\times(-{\pi\over 2}, {\pi \over 2})$. 
Combining \eqref{s5} and \eqref{s6}, we obtain
\begin{eqnarray}
&&  \int_0^{D}\int_{\S^{n-1}}{d\Omega_1\over |re_1-r_1\Omega_1|^m}dr\le{c(n)\over r^{m-2}}\int_0^{D}\int_{-{\pi \over 2}}^{\pi \over 2}{\cos(\theta_1)^{n-m}d\theta_1\over 
(r^2+r_1^2-2rr_1\sin(\theta_1))}dr_1\nonumber\\
&&\le{c(n)\over r^{m-2}}\int_0^{D}\int_{-{\pi \over 2}}^{\pi \over 2}{\cos(\theta_1)d\theta_1\over (r^2+r_1^2-2rr_1\sin(\theta_1))}dr_1
={c(n)\over r^{m-1}}\int_0^{D}{\ln\left({r+r_1\over r-r_1}\right)\over r_1}dr_1\nonumber\\
&&={c(n)\over r^{m-1}}\int_0^{D\over r}{\ln\left({1+\eta\over 1-\eta}\right)\over \eta}d\eta
\label{s7}
\end{eqnarray}
(we perform the change of variables $r_1=r\eta$, $dr_1=rd\eta$). Finally \eqref{s3} follows from \eqref{s7} and \eqref{s4} and the estimate 
$\int_0^{+\infty}{\ln\left({1+\eta\over 1-\eta}\right)\over \eta}d\eta<+\infty$.

We are now ready to prove \eqref{eq:norm_gamman+1}. Let $n\ge 2$. From \eqref{eq:gamma}, it follows that 
\begin{equation}
|\gamma_{n+1}(z_0,z_{n+1},v_{n+1})|\le\|\sigma_a\|_\infty\|k\|_{L^\infty(X\times\S^{n-1}\times\S^{n-1})}^{n+1}R(z_0,z_{n+1}),\label{s1}
\end{equation}
for a.e. $(z_0,z_{n+1},v_{n+1})\in X\times X\times\S^{n-1}$, where
\begin{equation}
R(z_0,z_{n+1})=\int_{X^n}{1\over\Pi_{i=1}^{n+1}|z_i-z_{i-1}|^{n-1}}dz_1\ldots dz_n,\label{s2}
\end{equation}
for $(z_0,z_{n+1})\in X\times X$.

Assume $n=2$. Combining \eqref{s3b} and \eqref{s2} we obtain
\begin{equation}
R(z_0,z_3)\le \int_X{C-C'\ln(|z_3-z_1|)\over|z_1-z_0|}dz_1. \label{s8.b}
\end{equation}
Performing the change of variables $z_1=z_0+r_1\Omega_1$, on the right hand side of \eqref{s8.b}, we obtain
\begin{equation}
R(z_0,z_3)\le \int_0^{D}\int_{\S^1}(C-C'\ln(| r_1 \Omega_1+z_0-z_3|))d \Omega_1d r_1. \label{s9.b}
\end{equation}

Assume $n\ge 3$.
Using \eqref{s2} and \eqref{s3}, we obtain 
\begin{equation}
R(z_0,z_{n+1})\le C\int_{X^2}{1\over |z_{n+1}-z_n|^{n-1}|z_n-z_{n-1}|^{n-1}|z_{n-1}-z_0|}dz_ndz_{n-1},\label{s80}
\end{equation}
where $C$ does not depend on $(z_0,z_{n+1})$.
Performing the change of variables $z_i=z_{i+1}+r_i\Omega_i$, $(r_i,\Omega_i)\in (0,+\infty)\times \S^{n-1}$, $i=n-1,n$, we obtain 
\begin{eqnarray}
\hspace{-1.5cm}R(z_0,z_{n+1})&\le&\int_{(0,D)^2\times(\S^{n-1})^2}{dr_{n-1}dr_nd\Omega_{n-1}d\Omega_n\over |z_{n+1}-z_0+r_n\Omega_n+r_{n-1}\Omega_{n-1}|}\nonumber\\
\hspace{-1.5cm}&=&\int_{(0,D)\times\S^{n-1}}\int_{(0,D)\times\S^{n-1}}{dr_{n-1}d\Omega_{n-1}\over ||z_{n+1}-z_0+r_n\Omega_n|e_1+r_{n-1}|}dr_nd\Omega_n,\label{s81}
\end{eqnarray}
where $e_1=(1,0,\ldots,0)\in \R^n$.
Performing the change of variables $\Omega=(\sin(\theta), \cos(\theta)\Theta)$, $(\theta,\Theta)\in (-{\pi\over 2},{\pi\over 2})\times \S^{n-2}$, we obtain
\begin{equation}
\int_0^{D}\int_{\S^{n-1}}{d\Omega\over ||w|e_1-r\Omega|}dr\le c(n)S(|w|),\label{s5**}
\end{equation}
for $w\in \R^n$, $w\not= 0$,  
where $c(n):=|\S^{n-2}|$ and $S(|w|)$ is defined by \eqref{s77}. 
Therefore from \eqref{s5**}, \eqref{s7.b} and \eqref{s81}, it follows that
\begin{equation}
R(z_0,z_{n+1})\le \int_{(0,D)\times\S^{n-1}}(C-C'\ln (|z_{n+1}-z_0+r_n\Omega_n|))dr_nd\Omega_n,\label{s100}
\end{equation}
where the positive constants $C$, $C'$ do not depend on $(z_0,z_{n+1})$.

Finally from \eqref{s9.b} and \eqref{s100}, it follows that
\begin{equation}
R(z_0,z_{n+1})\le \int_{(0,D)\times\S^{n-1}}(C-C'\ln (|z_{n+1}-z_0+r_n\Omega_n|))dr_nd\Omega_n,\label{s101}
\end{equation}
for $n\ge 2$ and for $(z_0,z_{n+1})\in X^2$, $z_0\not=z_{n+1}$, where the positive constants $C$, $C'$ do not depend on $(z_0,z_{n+1})$.

Let $n\ge 2$ and let $(z_0,z_{n+1})\in X^2$, $z_0\not=z_{n+1}$. From \eqref{s101} and the estimate
$|r_n+(z_0-z_{n+1})\cdot \Omega_n|\le |r_n\Omega_n+z_0-z_{n+1}|$ it follows that
\begin{eqnarray}
&&\hspace{-1cm} R(z_0,z_{n+1})\le c(n) CD-C'\int_{\S^{n-1}}\int_0^D\ln(| r_n +(z_0-z_{n+1})\cdot \Omega_n|)d r_nd\Omega_n\nonumber\\
&&\hspace{-1cm} \le c(n) (CD- C' \int_{-2D}^{2D}\ln(| r_n|)d r_n)= c(n) (CD-2C'\int_0^{2D}\ln(r_n)d r_n),\label{s11.b}
\end{eqnarray}
where $c(n):=|\S^{n-1}|$.
Statement \eqref{eq:norm_gamman+1} follows from \eqref{s1} and \eqref{s11.b}.

We prove \eqref{dec_noyau_stab}. We first obtain
\begin{equation}
\bar K^{n+1}(I-K)^{-1}J\phi(x)=\int_{X\times\S^{n-1}}\gamma_{n+1}(x,x',v')(I-K)^{-1}J\phi(x,v)dv,\label{t1b}
\end{equation}
for a.e. $x\in  X$ and for $\phi\in L^1(\Gamma_-,d\xi)$.
Therefore using \eqref{eq:norm_gamman+1} we obtain
\begin{equation}
\|\bar K^{n+1}(I-K)^{-1}J\phi(x)\|_{L^\infty(X)}\le C\|\phi\|_{L^1(\Gamma_-,d\xi)},\label{t2bb}
\end{equation}
for $\phi\in L^1(\Gamma_-,d\xi)$, where
$C:=\|\gamma_{n+1}\|_{L^\infty(X\times
  X\times\S^{n-1})}\|(I-K)^{-1}J\|_{{\cal L}(L^1(\Gamma_-,d\xi),
  L^1(X\times\S^{n-1}))}$.  Therefore there exists a (unique) function
$\Phi\in L^\infty(X\times \Gamma_-)$ such that
\begin{equation}
\bar K^{n+1}(I-K)^{-1}J\phi(x)=\int_{\Gamma_-}\Phi(x,x',v')
   \phi(x',v')d\xi(x',v').\label{t3b}
\end{equation}
Set $\Gamma_{n+1}(x,x',v'):=|\nu(x')\cdot v'|\Phi(x,x',v')$ for a.e. $(x,x',v')\in X\times\Gamma_-$ and recall the definition of $d\xi$. Then \eqref{dec_noyau_stab} follows 
from \eqref{t3b}.$\square$

\section{Derivation of stability estimates}
\label{sec:inverse}

\subsection{Estimates for ballistic part}
\label{sec:ballistic}

\paragraph{Proof of Theorem \ref{thm:ballistic}.}

Let $\phi\in L^\infty(X),\;\lVert\phi\rVert_{L^\infty(X)}\leq 1$ and $\psi\in L^1(\Gamma_-,d\xi)$, $\|\psi\|_{L^1(\Gamma_-,d\xi)}\leq 1$. We have :
\begin{equation}
\left|   \int_X\phi(x)\left[\left( A- \widetilde{ A}\right)\psi\right](x)dx\right|
\leq\lVert A- \widetilde{ A}\rVert_{ \mathcal{L}\left(L^1(\Gamma_-,d\xi),L^1(X)\right)}.\label{a1}
\end{equation}
Using \eqref{a1} and the decomposition of the albedo operator (see Theorem \ref{thm:decomp}) we obtain 
\begin{equation} 
\left|\Delta_0(\phi,\psi) \right|\leq \lVert  A- \widetilde{ A}\rVert_{ \mathcal{L}\left(L^1(\Gamma_-,d\xi),L^1(X)\right)}+ \left| \Delta_1(\phi,\psi) \right|,
\label{a2}
\end{equation}
where
\begin{eqnarray} 
\hspace{-1.5cm}\Delta_0(\phi,\psi)&=&\dint_{X\times\S^{n-1}} \phi(x)\left(\sigma_a(x,v')E(x,x-\tau_-(x,v')v')\right.\nonumber\\
\hspace{-1.5cm}&&\left.-\widetilde{\sigma}_a(x,v')\tilde E(x,x-\tau_-(x,v')v')\right)\psi(x-\tau_-(x,v')v',v')dv'dx,\label{a3}\\
\hspace{-1.5cm}\Delta_1(\phi,\psi)&=& \dint_X \phi(x)\dint_{\Gamma_-}
(\Gamma_1-\tilde \Gamma_1)(x,x',v')\psi(x',v')d\mu(x')dv'dx,\label{a4}
\end{eqnarray}
and where $\Gamma_1=\alpha_1+\Gamma_2$ (and $\tilde \Gamma_1=\tilde \alpha_1+\tilde \Gamma_2$).

Note that performing the change of variables $x=x'+tv'$ on the right-hand side of \eqref{a3}, we obtain
\begin{equation}
\Delta_0(\phi,\psi)=\dint_{\Gamma_-} \int_0^{\tau_+(x',v')}\phi(x'+tv')(\eta-\tilde \eta)(t;x',v')dt\psi(x',v')d\xi(x',v'),\label{a8}
\end{equation}
where
\begin{equation}
  \eta(t;x',v')=\sigma_a(x'+tv',v')e^{-  \int_0^t\sigma(x'+sv',v')ds},\  \tilde\eta(t;x',v')=\tilde\sigma_a(x'+tv',v')e^{-  \int_0^t\tilde\sigma(x'+sv',v')ds}.\label{a30}
\end{equation}

We use the following result: For any function $G\in L^1(\Gamma_-,d\xi)$ and for a.e. $(x_0',v_0')\in \Gamma_-$ there exists a sequence of functions $\psi_{\ep,x_0',v_0'}\in L^1(\Gamma_-,d\xi)$ (which does not depend on the function $G$), 
$\|\psi_{\ep,x_0',v_0'}\|_{L^1(\Gamma_-,d\xi)}=1$, $\psi_{\ep,x_0',v_0'}\ge 0$ and ${\rm supp}\psi_{\ep,x_0',v_0'}\subseteq \{(x',v')\in \Gamma_-\ |\ |x'-x'_0|+|v-v_0'|<\ep\}$ such that the following limit holds
\begin{equation}
\int_{\Gamma_-}G(x',v')\psi_{\ep,x_0',v_0'}(x',v')d\xi(x',v')\to G(x_0',v_0'), \textrm{ as }\ep\to 0^+\label{a6}
\end{equation}
(we refer the reader to \cite[Corollary 4.2]{McDST-IP-09} for the proof of this statement).  In particular, the limits \eqref{a6} holds for some sequence $\psi_{\ep,x_0',v_0'}$ when $(x_0',v_0')$ belongs to the Lebesgue set of $G$ denoted by ${\cal L}(G)$ and the complement of ${\cal L}(G)$ is a negligible subset of $\Gamma_-$ (see \cite{McDST-IP-09}). 

Let ${\cal L}:=\cap_{m\in \N\cup\{0\}} {\cal L}(G_m)$ where $G_m$ is the measurable function on $\Gamma_-$ defined by
\begin{equation}
G_m(x',v')=\int_0^{\tau_+(x',v')}t^m(\eta-\tilde\eta)(t;x',v')dt,\label{a5}
\end{equation}
for $m\in \N\cup\{0\}$. Note that the complement of ${\cal L}$ is still negligible.
Let $(x_0',v_0')\in {\cal L}$ and let $\phi\in C_0(0,\tau_+(x_0',v_0'))$ where  $C_0(0,\tau_+(x_0',v_0'))$ denotes the set of continuous and compactly supported functions on 
$(0,\tau_+(x_0',v_0'))$. Consider the sequence  $(\phi_m)\in (L^\infty(X))^\N$ defined by 
\begin{equation}
\phi_m(x)= \chi_{[0,{1\over m+1})}(|x'|)\phi(t),\label{a7}
\end{equation} 
for $x\in X$ and $m\in \N$ where $x=x_0'+tv_0'+x'$, $x'\cdot v_0'=0$,
where $\chi_{[0,{1\over m+1})}(t)=0$ when $t\ge {1\over m+1}$ and
$\chi_{[0,{1\over m+1})}(t)=1$ otherwise. The support of $\phi_m$
concentrates around the line which passes through $x_0'$ with
direction $v_0'$. From \eqref{a8}, \eqref{a6}, \eqref{a5} (and the
Stone-Weierstrass theorem), we obtain that
\begin{equation}
\lim_{\ep\to 0^+}\Delta_0(\phi_m,\psi_{\ep,x_0',v_0'})= \int_0^{\tau_+(x_0',v_0')}\phi_m(x_0'+tv_0')(\eta-\tilde\eta)(t;x_0',v_0')dt.\label{a8b} 
\end{equation}
For the single-scattering part, using \eqref{eq:Gamma2} and \eqref{eq:norm_simple} and \eqref{eq:norm_alpham} for $m=1$ we obtain:
\begin{equation}
|\Delta_1(\phi_m,\psi_{\ep,x_0',v_0'})|\le C\dint_{\Gamma_-}\Phi_m(x',v')\psi_{\ep,x_0',v_0'}(x',v')d\xi(x',v')\label{a20}
\end{equation}
where $C=\|{(\Gamma_1-\tilde\Gamma_1)(x,x',v')\over |\nu(x')\cdot v'|w_n(x,x',v')}\|_{\infty}$ and $\Phi$ is the function from $\Gamma_-$ to $\R$ defined by 
\begin{equation}
\Phi_m(x',v')=\int_Xw_n(x,x',v')|\phi_m(x)|dx\label{a21}
\end{equation}
(where $w_n(x,x',v')$ is defined in Theorem \ref{thm:stabscat}). From the definition of $w_n$, it follows that $\Phi_m$ is a bounded and continuous function on $\Gamma_-$. Therefore using Lebesgue dominated convergence theorem, we obtain
\begin{equation}
\lim_{\ep\to 0^+}\int_{\Gamma_-}\Phi_m(x',v')\psi_{\ep,x_0',v_0'}(x',v')d\xi(x',v')=\Phi_m(x_0',v_0')=\int_Xw_n(x,x_0',v_0')|\phi_m(x)|dx.\label{a22}
\end{equation}
Then using \eqref{a7}, \eqref{a22} and Lebesgue convergence theorem, we obtain

\noindent $\lim_{m\to +\infty}\int_Xw_n(x,x_0',v_0')|\phi_m(x)|dx=0$.
Therefore taking account of \eqref{a20} and \eqref{a22}, we obtain
\begin{equation}
\lim_{m\to +\infty} \lim_{\ep\to 0^+}\Delta_1(\phi_m,\psi_{\ep,x_0',v_0'})=0.\label{a24}
\end{equation}

Combining \eqref{a2} (with $\psi=\psi_{\ep,x',v'}$), \eqref{a8b}, \eqref{a7}, \eqref{a24}, we obtain
\begin{equation}
\left|\int_0^{\tau_+(x',v')}\phi(t)(\eta-\tilde\eta)(t;x',v')dt\right|\le \lVert  \A- \widetilde{ \A}\rVert_{ \mathcal{L}\left(L^1(\Gamma_-,d\xi),L^1(X)\right)},\label{a23}
\end{equation}
for $(x',v')\in {\cal L}$ and $\phi\in C_0(0,\tau_+(x',v'))$. From the density of $C_0(0,\tau_+(x',v'))$ in $L^1(0,\tau_+(x',v'))$, it follows that \eqref{a23} also holds for $\phi \in L^1(0,\tau_+(x',v'))$. Applying \eqref{a23} on
$\phi(t):=
\mathrm{sign}(\eta-\tilde\eta)(t;x',v')$
for a.e. $t\in (0,\tau_+(x',v'))$ (where $\mathrm{sign}(s)=1$ when $s\ge 0$ and $\mathrm{sign}(s)=-1$ otherwise), we obtain \eqref{eq:stabballistic}.
\hfill$\Box$

\paragraph{Proof of Theorem \ref{thm:stabscatfree}.}

Let us define
\begin{displaymath}
  \zeta(t):= \zeta(t;x',v') = e^{-\int_0^t \sigma_a(x'+sv',v')ds},\quad
  \eta(t):= \eta(t;x',v')=\sigma_a(x'+tv',v')\zeta(t;x',v').
\end{displaymath}
Note that $\eta(t)=\dr{}t\zeta(t)$.
Then,
\begin{displaymath}
\begin{array}{ll}
   |\zeta(t)-\tilde\zeta(t)| =
   \Big|\dint_0^t \dr{}t(\zeta-\tilde\zeta)(s)ds  
   \Big|
  \leq \dint_0^t |\eta-\tilde\eta|(s)ds \leq 
  \|A-\tilde A\|_{{\mathcal L}(L^1(\Gamma_-,d\xi);L^1(X))},
\end{array}
\end{displaymath}
thanks to \eqref{eq:stabballistic}. The point-wise (in $t$) control on
$\zeta(t)$ and the estimate \eqref{eq:stabballistic} for $\eta(t)$
show by application of the triangle inequality and the fact that
$\zeta(t)$ is bounded from below by the positive constant $e^{-M{\rm
    diam}(X)}$ that
\begin{equation}
  \label{eq:ineqsigmaa}
  \begin{array}{lll}
     \dint_0^{\tau_+(x',v')} &\Big| \sigma_a(x'+tv',v') -
      \tilde\sigma_a(x'+tv',v')\Big| dt
      \leq  C\|A-\tilde A\|_{{\mathcal L}(L^1(\Gamma_-,d\xi);L^1(X))}.
    \end{array}
\end{equation}
Here, the constant $C$ depends on $M$ and is of order $e^{M{\rm
    diam}(X)}$. Not surprisingly, reconstructions deteriorate when the
optical depth $M{\rm diam}(X)$ of the domain increases. This shows
that
\begin{displaymath}
  \dint_X   \big|\sigma_a(x,v')- 
 \tilde\sigma_a(x,v')\big| dx\leq 
  C\|A-\tilde A\|_{{\mathcal L}(L^1(\Gamma_-,d\xi);L^1(X))}.
\end{displaymath}
This concludes the proof of the theorem. $\square$

\paragraph{Proof of Theorem \ref{thm:stabsym}.}
Let ${\cal G}:=\{(t,x',v')\in (0,+\infty)\times\Gamma_-\ |\ t\in (0,\tau_+(x',v'))\}$ and let $h\in L^{\infty}({\cal G})$ be defined by
\begin{equation}
 h(t;x',v')=-\dint_0^t\sigma(x'+sv',v')ds+\dint_t^{\tau_+(y',v')}\sigma(y'+sv',v')ds,\label{b1}
\end{equation}
for $(t,x',v')\in {\cal G}$. The function $\tilde h\in L^\infty({\cal G})$ is defined similarly. 

We first prove a stability estimate \eqref{eq:hStab} on $h,$ $\tilde h$.
Let $(y_0,v_0)\in \Gamma_-$, $t\in (0,\tau_+(y_0,v_0))$. Set $(y_1,v_1):=(y_0+\tau_+(y_0,v_0)v_0,-v_0)$ and $t_1=\tau_+(y_0,v_0)-t$.
Due to the symmetry of  $\sigma_a$, $\sigma$ with respect to the speed variable $v$ ($\sigma_a$, $\widetilde{\sigma}_a$ bounded from below), we obtain:
\begin{equation}
\frac{\eta(t;y_0,v_0)}{\eta(t_1;y_1,v_1)}=e^{h(t;y_0,v_0)},\ 
\frac{ \tilde \eta(t;y_0,v_0)}{ \tilde \eta(t_1; y_1,v_1)}=
e^{ \widetilde{h}(t;y_0,v_0)},\label{b2}
\end{equation}
where $\eta$ is defined by \eqref{a30}. 
Note that from \eqref{b1}, it follows that $|h(t;y_0,v_0)|\le \|\sigma\|_{\infty}\tau_+(y_0,v_0)\le D\|\sigma\|_{\infty}$ where $D$ denotes the diameter of $X$.
A similar estimate is valid for $\tilde h$ and $\tilde \sigma$.
Therefore  using the fact that $|a-\tilde a|\leq e^{-\min(a,\tilde a)}|e^a-e^{\tilde a}|$ for $\tilde a=h(t)$, we obtain
\begin{equation}
|h- \widetilde{h}|(t;y_0,v_0)
\leq e^{D\max(\|\sigma\|_{\infty},\|\tilde\sigma\|_\infty)}\left|\frac{\eta(t;y_0,v_0)}{\eta(t_1;y_1,v_1)}-\frac{ \widetilde{\eta}(t;y_0,v_0)}{ \widetilde{\eta}(t_1;y_1,v_1)}\right|.
\label{b6}
\end{equation}

Note that from \eqref{a30} it follows that
\begin{equation}
 0<\sigma_0e^{-D\lVert\sigma\rVert_{\infty}}\leq \eta(s;y,v)\leq\lVert\sigma_a\rVert_{\infty}.\label{b4}
\end{equation}
for $(y,v)\in \Gamma_-$ and $s\in (0,\tau_+(y,v))$, where $D$ is the
diameter of $X$. A similar estimate is valid for $(\tilde
\sigma_a,\tilde \eta)$.  Therefore using the equality ${a\over
  b}-{\tilde a\over \tilde b}={(a-\tilde a)\tilde b+(b-\tilde b)\tilde
  a\over b\tilde b}$ (for $a={\eta(t;y_0,v_0)}$,
$b={\eta(t_1;y_1,v_1)}$), we obtain
\begin{eqnarray}
\hspace{-1cm}&&\left|\frac{\eta(t;y_0,v_0)}{\eta(t_1;y_1,v_1)}-\frac{ \widetilde{\eta}(t;y_0,v_0)}{ \widetilde{\eta}(t;y_1,v_1)}\right|\nonumber\\
\hspace{-1cm}&\leq& \frac{e^{D\left(\lVert \sigma\rVert_{\infty}+\lVert  \widetilde\sigma\rVert_{\infty}\right)}}{\sigma_0 \widetilde{\sigma_0}}\Big(
\eta(t;y_0,v_0)|\eta-\tilde \eta|(t_1;y_1,v_1)\nonumber
+\eta(t_1;y_1,v_1)|\eta-\tilde\eta|(t;y_0,v_0)\Big)\nonumber\\
\hspace{-1cm}&\leq& \frac{e^{D\left(\lVert \sigma\rVert_{\infty}+\lVert  \widetilde\sigma\rVert_{\infty}\right)}}{\sigma_0 \widetilde{\sigma_0}}
\Big(\lVert\sigma_a\rVert_{\infty}|\eta-\tilde\eta|(t_1;y_1,v_1)
+\lVert \widetilde\sigma_a\rVert_{\infty}|\eta-\tilde\eta|(t;y_0,v_0)\Big).\label{b3}
\end{eqnarray}
Using  \eqref{b3}, integrating in the $t$ variable ($t_1=\tau_+(y_0,v_0)-t$) and using \eqref{eq:stabballistic}, we obtain
\begin{eqnarray}
\hspace{-1cm}\int_0^{\tau_+(y_0,v_0)}\left|\frac{\eta(t;y_0,v_0)}{\eta(t_1;y_1,v_1)}-\frac{ \widetilde{\eta}(t;y_0,v_0)}{ \widetilde{\eta}(t_1;y_1,v_1)}\right|dt
&\leq& 
\frac{e^{D\left(\lVert \sigma\rVert_{\infty}+\lVert  \widetilde\sigma\rVert_{\infty}\right)}}{\sigma_0 \widetilde{\sigma_0}}\left(
\lVert\sigma_a\rVert_{\infty }
+
\lVert \widetilde\sigma_a\rVert_{\infty }
\right)\nonumber\\
\hspace{-1cm}&&\times\lVert\A- \widetilde{\A}\rVert_{\mathcal{L}\left(L^1(\Gamma_-,d\xi),L^1(X)\right)}.\label{b5}
\end{eqnarray}

Combining \eqref{b6} and \eqref{b5} we obtain
\begin{equation}
\label{eq:hStab}
\int_0^{\tau_+(y_0,v_0)}|h- \widetilde{h}|(t;y_0,v_0)dt
\leq C \lVert\A- \widetilde{\A}\rVert_{\mathcal{L}\left(L^1(\Gamma_-,d\xi),L^1(X)\right)}
\end{equation}
for a.e. $(y_0,v_0)\in \Gamma_-$, where the constant $C$ depends only on $D$, $\|\sigma\|_\infty$, $\|\tilde \sigma\|_\infty$, $\|\sigma_a\|_\infty$ and $\|\tilde\sigma_a\|_\infty$. 
We now prove \eqref{b20}.
Estimate \eqref{b20} is, in particular, a consequence of the identities
$$
\frac{dh(t; y_0,v_0)}{dt}=-2\sigma(y_0+sv_0,v_0),\ 
\frac{d \widetilde{h}(t; y_0,v_0)}{dt}=-2 \widetilde\sigma(y_0+sv_0,v_0),
$$
for a.e. $(y_0,v_0)\in \Gamma_-$ and $t\in (0,\tau_+(y_0,v_0))$.

Consider the operator $T_{\delta\sigma(.,v)}$ associated with $\delta\sigma=\sigma- \widetilde\sigma$ :
$$
T_{\delta\sigma(.,v)}:W^{1,\infty}_0(X)\to \R,\ \phi\mapsto\int_X\delta\sigma(x,v)\varphi(x)dx 
$$
for $v\in \S^{n-1}$. Using the change of variable $x=y+tv$
($dx=|\nu(y)\cdot v|dtdy$, $y\in \partial X_-(v):=\{x'\in \pa X\ |\ 
\nu(x')\cdot v<0\}$, $t\in (0,\tau_+(y,v))$) we have
$$
\left|\langle T_{\delta\sigma(.,v)},\varphi\rangle\right|=\left|\int_{\partial X_-(v)}|\nu(y)\cdot v|\int_0^{\tau_+(y,v)}\delta\sigma(y+tv,v)\varphi(y+tv)dtd\mu(y) \right|,
$$
and using integration by part in the inner integral we have:
\begin{equation}
\left|\langle T_{\delta\sigma(.,v)},\varphi\rangle\right|
\leq\dint_{\partial X_-(v)}{|\nu(y)\cdot v|\over 2}\left|\dint_0^{\tau_+(y,v)}\!\!\!(h-\tilde h)(t;y,v)
v\cdot\nabla\varphi(y+tv)dt\right|d\mu(y) \label{cc1}
\end{equation}
\begin{eqnarray}
&\leq&
\frac 1 2
\lVert\nabla\varphi\rVert_{\infty}
\dint_{\partial X_-(v)}\dint_0^{\tau_+(y,v)}\left|h-\widetilde{h}\right|(t;y,v)dt
d\mu(y)\nonumber\\
&\leq&
C\lVert\A- \widetilde{\A}\rVert_{\mathcal{L}\left(L^1(\Gamma_-,d\xi),L^1(X)\right)}
\lVert \varphi\rVert_{W^{1,\infty}_0(X)},
\end{eqnarray}
for a.e. $v\in \S^{n-1}$ (we used \eqref{eq:hStab}).
Identifying $\delta \sigma$ and $T_{\delta\sigma}$, we obtain 
\begin{equation}
\lVert \sigma- \widetilde{\sigma}\rVert_{L^\infty\left( \S^{n-1},W^{-1,1}(X)\right)}\leq C\lVert\A- \widetilde{\A}\rVert_{\mathcal{L}\left(L^1(\Gamma_-,d\xi),L^1(X)\right)}.\label{b20}
\end{equation}

We prove the following estimate
\begin{equation}
\|\sigma_a-\tilde\sigma_a\|_{L^\infty(\S^{n-1},L^1(X))}\le C\|A-\tilde A\|_{{\cal L}(L^1(\Gamma_-);L^1(X))},\label{b21}
\end{equation} 
where $C$ depends on the uniform bounds $\sigma_0$ and $M$.
Combining \eqref{b20} and \eqref{b21}, we obtain \eqref{eq:stabsym}, which completes the proof of Theorem \ref{thm:stabsym}. 
The estimate on $\sigma_a(x,v')$ comes from an estimate on
$\int_0^t \sigma(x'+sv',v')ds$ by the triangle inequality. The latter is
given by
\begin{displaymath}
  2\int_0^t \sigma(x'+sv',v')ds = 
  \int_0^{\tau_+(x',v')} \hspace{-1cm} \sigma(x'+sv',v')ds 
   + \Big(\int_0^t-\int_t^{\tau_+(x',v')}\Big)\sigma(x'+sv',v')ds.
\end{displaymath}
We just obtained control of 
\begin{displaymath}
  \Big(\int_0^t-\int_t^{\tau_+(x',v')}\Big)(\sigma-\tilde\sigma)(x'+sv',v')ds,
\end{displaymath}
in the $L^1$ sense in the $t-$variable. It thus remains to control the
constant term $\int_0^{\tau_+(x',v')}
(\sigma-\tilde\sigma)(x'+sv',v')ds$.

Note that the latter term is nothing but the X-ray transform (Radon
transform when $n=2$) of $\sigma$ along the line of direction $v'$
passing through $x'$. In the setting of measurements that are supposed
to be accurate in ${\mathcal L}(L^1(\Gamma_-,d\xi);L^1(X))$, the line
integral is not directly captured as it corresponds to a measurement
performed at a point $x=x'+\tau_+(x',v')v'$. This is the reason why we
assume that $\sigma$ is known in the $\delta-$vicinity of $\partial
X$.  

Knowledge of $\sigma$ and $\tilde\sigma$ in the $\delta_0$-vicinity of
$\partial X$ allows one to control $\int_0^{\tau_+(x',v')}
\sigma(x'+sv',v')ds$ by $\|A-\tilde A\|_{{\mathcal
    L}(L^1(\Gamma_-,d\xi);L^1(X))}$. When the X-ray transform of $\sigma$
is well captured by available measurements, as for instance in the
presence of boundary measurements \cite{B-IP-09}, then $\delta_0$ can
be set to $0$.

More precisely, we find that 
\begin{displaymath}
  \int_0^{\tau_+(x',v')}
(\sigma-\tilde\sigma)(x'+sv',v')ds =
  \int_0^{\tau_+(x',v')} \phi(s)
(\sigma-\tilde\sigma)(x'+sv',v')ds,
\end{displaymath}
where $\phi(s)\in C^\infty_0(0,\tau_+(x',v'))$ is equal to $1$ for
$\delta_0<s<\tau_+(x',v')-\delta_0$. Integrating by parts, this shows that 
\begin{displaymath}
  \begin{array}{ll}
  \quad  \Big|\dint_0^{\tau_+(x',v')}
(\sigma-\tilde\sigma)(x'+sv',v')ds \Big|
 \\\leq\!  
  \|(\sigma-\tilde\sigma)(x'+tv',v')\|_{W_t^{-1,1}(0,\tau_+(x',v'))}
  \|\phi\|_{W^{1,\infty}(0,\tau_+(x',v'))} 
  \leq\! C \|A-\tilde A\|_{{\mathcal
    L}(L^1(\Gamma_-,d\xi);L^1(X))},
   \end{array}
\end{displaymath}
thanks to estimate \eqref{eq:hStab}. 
By Lipschitz regularity of the exponential, we thus have that
\begin{equation}
  \dint_0^{\tau_+(x',v')}
   \Big| e^{-\int_0^t\sigma(x'+sv',v')}-
    e^{-\int_0^t\tilde\sigma(x'+sv',v')}\Big| dt \leq 
    C \|A-\tilde A\|_{{\mathcal L}(L^1(\Gamma_-,d\xi);L^1(X))}.\label{ee1}
\end{equation}
The stability result \eqref{b21} on $\sigma_a(x,v')$ follows from
\eqref{eq:stabballistic} and the triangle inequality as in the proof
of Theorem \ref{thm:stabscatfree}. $\square$

\paragraph{Proof of Corollary \ref{cor:regsigma}.}
Let $p>1$. We first derive \eqref{eq:postbdsigma} from  the following estimate
\begin{equation}
\lVert \sigma- \widetilde{\sigma}\rVert_{L^\infty\left( \S^{n-1},W^{-1,p}(X)\right)}\leq C\lVert\A- \widetilde{\A}\rVert_{\mathcal{L}\left(L^1(\Gamma_-,d\xi),L^1(X)\right)}^{\frac 1 p}.
\label{c1}
\end{equation}
Assuming that $\sigma-\widetilde{\sigma}$ is bounded by $C_0$ in
$L^\infty\left( \S^{n-1},W^{r,p}(X)\right)$, then using \eqref{c1} and using the complex
interpolation result \cite[Theorem 6.4.5 pp. 153]{bergh-lofstrom}, we obtain
\begin{equation}
\|\sigma-\tilde \sigma\|_{W^{s,p}(X)}\le \|\sigma-\tilde \sigma\|_{W^{-1,p}(X)}^{1-\theta}\|\sigma-\tilde \sigma\|_{W^{r,p}(X)}^\theta\le  C'\lVert\A- \widetilde{\A}\rVert_{\mathcal{L}\left(L^1(\Gamma_-,d\xi),L^1(X)\right)}^{1-\theta\over p},\label{c2}
\end{equation}
for  $-1\le s\le r$, where $s=(1-\theta)\times (-1)+r\theta$ and $C'=C^{1-\theta}C_0^\theta$. This proves \eqref{eq:postbdsigma}.

We prove \eqref{c1}. From \eqref{b1}, it follows that $|h-\tilde h|(t;x',v')\le (|h|+|\tilde h|)(t;y_0,v_0)\le D(\|\sigma\|_{\infty}+\|\tilde \sigma\|_\infty)$, where $D$ is the diameter of $X$. Hence we obtain 
\begin{equation}
\left|h-\widetilde{h}\right|^p(t;x',v')\le D^{p-1}(\|\sigma\|_{\infty}+\|\tilde \sigma\|_\infty)^{p-1}|h-\tilde h|(t;x',v'),\label{c3}
\end{equation}
for $(x',v')\in\Gamma_-$ and $t\in (0,\tau_+(x',v'))$.  Using
\eqref{cc1} and H\"{o}lder inequality, and using \eqref{c3}, we obtain
\begin{eqnarray}
\hspace{-2cm}\left|\int_X(\sigma-\tilde \sigma)(x,v)\varphi(x)dx\right|&\leq&
 \frac{1}{2}\left(\dint_{\partial X_-(v)}\int_0^{\tau_+(y,v)}\!\!\!|h-\widetilde{h}|^p(t;y,v)dtd\mu(y)\right)^{\frac 1 p} \lVert \nabla\varphi\rVert_{L^{p'}(X)}\nonumber\\
\hspace{-2cm}
&&\leq C\lVert\A- \widetilde{\A}\rVert_{\mathcal{L}\left(L^1(\Gamma_-,d\xi),L^1(X)\right)}^{\frac 1 p}\lVert \varphi\rVert_{W^{1,p'}(X)},\label{c4}
\end{eqnarray}
for a.e. $v\in \S^{n-1}$  and for $\varphi\in W_0^{1,p'}(X):=\{\psi\in L^{p'}(X)\ |\ {\rm supp}\psi\subset X,\ \nabla\psi\in L^{p'}(X,\C^n)\}$ where ${p'}^{-1}+p^{-1}=1$.
Estimate \eqref{c4} proves \eqref{c1}.$\square$

\subsection{Estimates for single scattering}
\label{sec:single}

\paragraph{Proof of Theorem \ref{thm:stabscat}.}
Let $(x,x')\in X\times \pa X$. Set $v'=\widehat{x-x'}$ and let ${v'}^\bot\in \S^{n-1}$ be such that $v'\cdot {v'}^\bot=0$. Let $t_0'=|x-x'|$, then $x=x'+t_0'v'$. First assume $n=2$.
Using the equality $\alpha_1=\Gamma_1-\Gamma_2$, and using \eqref{eq:Gamma2}, \eqref{eq:norm_simple}, we obtain
\begin{eqnarray}
&&\hspace{-1.5cm}{|\alpha_1-\tilde \alpha_1|(x+\ep {v'}^\bot,x',v')\over |\nu(x')\cdot v'|\left(1+\ln\left({|(t_0'-\tau_+(x',v'))v'+\ep {v'}^\bot|+(\tau_+(x',v')-t_0')
\over |t_0'v'+\ep{v'}^\bot|-t_0'}\right)\right)}\nonumber\\
&&\hspace{-1.5cm}\le 
\left\|{(\Gamma_1-\tilde \Gamma_1)(z,z',w')\over |\nu(z')\cdot w'|w_2(z,z',w')}
\right\|_{\infty}\!\!\!\!
+{\left\|{(\Gamma_2-\tilde \Gamma_2)(z,z',w')\over |\nu(z')\cdot w'|}\right\|_{\infty}\over 1+\ln\Big({|(t_0'-\tau_+(x',v'))v'+\ep {v'}^\bot|+(\tau_+(x',v')-t_0')
\over |t_0'v'+\ep{v'}^\bot|-t_0'}\Big)}\label{f1}
\end{eqnarray}
for $\ep >0$. Therefore using \eqref{f1} as $\ep\to 0^+$ and \eqref{eq:asyalpha1}, we obtain \eqref{eq:dim2}. 

Assume $n\ge 3$.
Using the equality $\alpha_1=\Gamma_1-\Gamma_2$, and using \eqref{eq:Gamma2}, \eqref{eq:norm_alpham} (for ``$m=1$''), we obtain
\begin{eqnarray}
\hspace{-1cm}&&\ep^{n-2}{|\alpha_1-\tilde \alpha_1|(x+\ep {v'}^\bot,x',v')\over|\nu(x')\cdot v'|}\le 
\left\|{(\Gamma_1-\tilde \Gamma_1)(z,z',w')\over |\nu(z')\cdot w'|w_n(z,z',w')}\right\|_{L^\infty(X\times\Gamma_-)}\nonumber\\
\hspace{-1cm}&&+\ep\left\|{|z-z'-((z-z')\cdot w')w'|^{n-3}(\Gamma_2-\tilde \Gamma_2)(z,z',w')\over |\nu(z')\cdot w'|}\right\|_{L^\infty(X\times\Gamma_-)}
\label{f2}
\end{eqnarray}
for $\ep>0$. Therefore using \eqref{f2} as $\ep\to 0^+$ and
\eqref{eq:asyalpha1}, we obtain \eqref{eq:dim3}.  {}\hfill $\Box$

\paragraph{Proof of  Lemma \ref{hgn=3}.}
First consider the case $n=2$. The statement in Lemma \ref{hgn=3} is a
straightforward consequence of \eqref{g6} and we have the following
inversion formula:
\begin{equation}
g=\Big({\pi h(g)-1\over \pi h(g)+1}\Big)^{\frac12},\ g\in [0,1).\label{d1}
\end{equation}
Let now $n=3$.  Using \eqref{g7} and the identity
$\cos(\theta)=1-2\sin^2({\theta\over 2})$, $\theta\in \R$, we obtain
$$
2\pi h(g)={1+g\over (1-g)^2}\int_0^\pi 
\dfrac12
{\left(1+{4g\over (1-g)^2}\sin^2\big({\theta\over 2}\big)\right)^{-{3\over 2}}}
d\theta,
$$
for $g\in [0,1)$.  Performing the change of variables
$\theta=2\arcsin({t\over \sqrt{1+t^2}})$ ($d\theta={2dt\over 1+t^2}$),
we obtain
$$
2\pi h(g)= {1+g\over (1-g)^2}\int_0^{+\infty} 
(1+t^2)^{\frac12}\bigg(1+\left({1+g\over 1-g}\right)^2t^2\bigg)^{-\frac32}dt.
$$
Performing the change of variables $v={1+g\over 1-g}t$, we obtain
\begin{eqnarray}
\hspace{-2cm}2 \pi h(g)&=&\int_0^{+\infty} \sqrt{\left({1\over 1-g}\right)^2+\left({1\over 1+g}\right)^2v^2\over \left(1+v^2\right)^3}dv
={1\over 1-g}\int_0^{+\infty} \sqrt{1+\left({1-g\over 1+g}\right)^2v^2\over \left(1+v^2\right)^3}dv,\label{h2}
\end{eqnarray}
for $g\in [0,1)$. Note that from the above, it follows that $2\pi
h(g)={1\over 1-g}+o({1\over 1-g})$ as $g\to 1^-$ (where we used the
integral value $\int_0^{+\infty}{dv\over (1+v^2)^{3\over
    2}}=\left[{v\over \sqrt{1+v^2}}\right]_0^{+\infty}=1$).

Differentiating \eqref{h2} with respect to $g$, we obtain
\begin{eqnarray}
\hspace{-2cm}
2\pi \dot h(g)&=&{1\over (1-g)^3}\int_0^{+\infty}{dv\over (1+v^2)^{3\over 2}\sqrt{\omega_1^2(g)+\omega_2^2(g)v^2}}
-{1\over (1+g)^3}h_2(g),\label{h4}
\end{eqnarray}
for $g\in [0,1)$, where $\dot h={d h\over dg}$, $\omega_1(g):={1\over
  1-g}$, $\omega_2(g):={1\over 1+g}$, and
\begin{equation}
h_2(g):=\int_0^{+\infty}{v^2dv \over (1+v^2)^{3\over 2}\sqrt{\omega_1^2(g)+\omega_2^2(g)v^2}}.\label{h5}
\end{equation}
Integrating by parts, we obtain
\begin{eqnarray}
h_2(g)
&=&\int_0^{+\infty}{\omega_1^2(g)dv\over (1+v^2)^{1\over 2}\left({\omega_1^2(g)+\omega_2^2(g)v^2}\right)^{3\over 2}},\label{h6}
\end{eqnarray}
for $g\in [0,1)$, where we use that a primitive of the function
$r(v):=v(1+v^2)^{-{3\over 2}}$ is given by the function
$R(v):=-(1+v^2)^{-{1\over 2}}$ and where we used that the derivative
of the function
$s(v):=\left(\omega_1^2(g)+\omega_2^2(g)v^2\right)^{-{1\over 2}}$ is
given by $\dot
s(v)=\omega_1^2(g)\left(\omega_1^2(g)+\omega_2^2(g)v^2\right)^{-{3\over
    2}}$.  Performing the change of variables ``$v$''$={1-g\over
  1+g}v$ on the right hand side of \eqref{h6}, we obtain
\begin{equation}
h_2(g)=\omega_2^2(g)\int_0^{+\infty}(1+v^2)^{-{3\over 2}}\left(1+\left({1+g\over 1-g}\right)^2v^2\right)^{-{1\over 2}}dv,\label{h7}
\end{equation}
for $g\in [0,1)$. Combining \eqref{h4} and \eqref{h7}, we obtain
\begin{eqnarray}
2\pi \dot h(g)&=&{1\over (1-g)^2}\Big[\int_0^{+\infty}{(1+v^2)^{-{3\over 2}}
\Big(1+\left({1-g\over 1+g}\right)^2v^2\Big)^{-\frac12}}dv\nonumber\\
&&-\left({1-g\over 1+g}\right)^2\int_0^{+\infty}{(1+v^2)^{-{3\over 2}}\Big({1+\left({1+g\over 1-g}\right)^2v^2}\Big)^{-{1\over 2}}}dv\Big],\label{h8}
\end{eqnarray}
for $g\in [0,1)$.
Using the estimate $1+\left({1+g\over 1-g}\right)^2v^2>1+\left({1-g\over 1+g}\right)^2v^2$ for $v\in (0,+\infty)$ and $g\in (0,1)$, we obtain that the second integral on the 
right hand side of \eqref{h8} is less than  the first integral on the right-hand side of \eqref{h8} for $g\in (0,1)$. Therefore using also that the second integral is 
multiplied by $\left({1-g\over 1+g}\right)^2(<1)$, we obtain $\dot h(g)>0$ for $g\in (0,1)$, which proves that $h$ is strictly increasing on $(0,1)$. $\square$

\paragraph{Proof of Theorem \ref{thm:stabHG}}
We prove \eqref{eq:HGdim2}. From \eqref{eq:dim2}, \eqref{eq:dim3}, \eqref{g1} and \eqref{g2}, it follows that
\begin{equation}
|E(x,x')\sigma_g(x)-\tilde E(x,x')\sigma_g(x)|\le \left\|{(\Gamma_1-\tilde\Gamma_1)(x,x',v')\over |\nu(x')\cdot v'|w_n(x,x',v')}\right\|_{\infty},\label{e1}
\end{equation}
for $(x,x')\in X\times\Gamma_-$.  Using \eqref{e1} and the estimate
$\min(E(x,x'), \tilde E(x,x'))\ge e^{-D\max(\|\sigma\|_\infty,\|\tilde
  \sigma\|_\infty)}$ (see \eqref{eq:E}) where $D$ is the diameter of
$X$, we obtain
\begin{eqnarray}
\hspace{-2cm}|\sigma_g-\tilde\sigma_g|(x)&\le& e^{D\max(\|\sigma\|_\infty,\|\tilde \sigma\|_\infty)}\Big(\left\|{(\Gamma_1-\tilde\Gamma_1)(x,x',v')\over |\nu(x')\cdot v'|w_n(x,x',v')}\right\|_{\infty}
+|E-\tilde E|(x,x')\Big),\label{e2}
\end{eqnarray}
for $(x,x')\in X\times\Gamma_-$ (we used the identity $ab-\tilde
a\tilde b=(a-\tilde a)b+(b-\tilde b)\tilde a$ for $a:=\sigma_g(x)$ and
$b:=E(x,x')$).  Let $(x',v')\in \Gamma_-$. Integrating \eqref{e2} over
the line which passes through $x'$ with direction $v'$ and using the
stability estimate \eqref{ee1} and \eqref{eq:E}, we obtain
\begin{eqnarray}
\hspace{-2.5cm}
\int_0^{\tau_+(x',v')}\hspace{-1cm}
|\sigma_g-\tilde\sigma_g|(x'\!+tv')dt&\le& 
C\Bigg\|{(\Gamma_1-\tilde\Gamma_1)(x,x',v')\over |\nu(x')\cdot v'|w_n(x,x',v')}\Bigg\|_{\infty}
\hspace{-0.4cm}
+C\|A-\tilde A\|_{{\cal L}(L^1(\Gamma_-,d\xi), L^1(X)) },\label{e2b}
\end{eqnarray}
where $C= e^{D\max(\|\sigma\|_\infty,\|\tilde
  \sigma\|_\infty)}\max(D,C)$ and $C$ is the constant on the
right-hand side of \eqref{ee1}.  Integrating \eqref{e2b} over $\pa
X_-(v'):=\{x'\in \pa X\ |\ v'\cdot \nu(x')<0\}$ with measure
$|\nu(x')\cdot v'|d\mu(x')$, we obtain \eqref{eq:HGdim2}.
We now prove \eqref{eq:HGdim23}. Assume $\|h(g)\|_{W^{1,\infty}(X)}\le \|h(\tilde g)\|_{W^{1,\infty}(X)}$.
Using \eqref{g5} and $\min(\sigma_s,\tilde\sigma_s)\ge \sigma_{s,0}$, we obtain
\begin{equation} 
\|h(g)- h(\tilde g)\|_{L^1(X)}\le {1\over \sigma_{s,0}}\Big(\|(\sigma_s-\tilde \sigma_s)h(g)\|_{L^1(X)}+\|\sigma_g-\tilde\sigma_g\|_{L^1(X)}\Big)\label{e3}
\end{equation}
(we used the identity $ab-\tilde a\tilde b=(a-\tilde a)b+(b-\tilde b)\tilde a$ for $a=\sigma_s$ and $b=h(g)$).
Using the identity $\sigma_s=\sigma-\sigma_a$, we have
\begin{eqnarray}
\hspace{-2cm}&& \|(\sigma_s-\tilde \sigma_s)h(g)\|_{L^1(X)}\le \|\sigma_a-\tilde\sigma_a\|_{L^1(X)}\|h(g)\|_{L^\infty(X)}+\|(\sigma-\tilde\sigma)h(g)\|_{L^1(X)}\nonumber\\
\hspace{-2cm}&\le &\|\sigma_a-\tilde\sigma_a\|_{L^1(X)}\|h(g)\|_{L^\infty(X)}+\|\sigma-\tilde\sigma\|_{W^{-1,1}(X)}\|\nabla h(g)\|_{L^\infty(X)}\label{e4}
\end{eqnarray}
(we used the fact that $\sigma=\tilde \sigma$ at the vicinity of the boundary $\pa X$).
Finally combining \eqref{e3}, \eqref{e4}, \eqref{eq:HGdim2} and \eqref{eq:stabsym}, we obtain \eqref{eq:HGdim23}.$\square$

\appendix
\section{}
For $m\ge 2$, let $\beta_m$ denotes the distributional kernel of the
operator $K^m$ where $K$ is defined by \eqref{eq:K}.  We first give
the explicit expression of $\beta_2$, $\beta_3$. Then by induction we
give the explicit expression of the kernel $\beta_m$. Finally we prove
Lemma \ref{lem:dec}.  \\From \eqref{eq:K} it follows that
\begin{eqnarray}
&&\hspace{-1cm}K^2\psi(x,v)=\int_0^{\tau_-(x,v)}\int_{\S^{n-1}}k(x-tv,v_1,v)\int_0^{\tau_-(x-tv,v_1)}\!\!\!\!\!\!\!\!\!\!\!\!\!E(x,x-tv,x-tv-t_1v_1)\nonumber\\
&&\hspace{-1cm}\times\int_{\S^{n-1}}k(x-tv-t_1v_1,v',v_1)\psi(x-tv-t_1v_1,v')dv'dt_1dv_1dt,
\label{t1}
\end{eqnarray}
for a.e. $(x,v)\in X\times\S^{n-1}$ and for $\psi\in L^1(X\times\S^{n-1})$.
Performing the change of variables $x'=x-tv-t_1v_1$ ($dx'=t_1^{n-1}dt_1dv_1$) on the right hand side of \eqref{t1}, we obtain
\begin{eqnarray}
K^2\psi(x,v)&=&\int_0^{\tau_-(x,v)}\int_Xk(x-tv,\widehat{x-tv-x'},v)E(x,x-tv,x')\nonumber\\
&&\times\int_{\S^{n-1}}k(x',v',\widehat{x-tv-x'})\psi(x',v')dv'dx'dt,
\label{t2b}
\end{eqnarray}
for a.e. $(x,v)\in X\times\S^{n-1}$ and for $\psi\in L^1(X\times\S^{n-1})$. 
Therefore
\begin{equation}
\beta_2(x,v,x',v')=\int_0^{\tau_-(x,v)}\!\!\!\!\!\!\!E(x,x-tv,x'){k(x-tv,v_1,v)k(x',v',v_1)_{|v_1= \widehat{x-tv-x'}}\over |x-tv-x'|^{n-1}}dt,\label{t2}
\end{equation}
for a.e. $(x,v,x',v')\in X\times\S^{n-1}\times X\times \S^{n-1}$. From
\eqref{t2} and \eqref{eq:K} it follows that
$$
K^3\psi(x,v)=\int_{(0,\tau_-(x,v))\times\S^{n-1}}\!\!\!\!\!\!\!\!\!\!\!\!\!\!\!\!\!\!\!\!\!\!\!\!\!\!\!\!\!\!E(x,x-tv)k(x-tv,v_1,v)\int_{X\times\S^{n-1}}\!\!\!\!\!\!\!\!\!\!\!\beta_2(x-tv,v_1,x',v')\psi(x',v')dx'dv'dv_1dt
$$
\begin{eqnarray}
&&\hspace{-2cm}=\int_{X\times\S^{n-1}}\hspace{-1cm}\psi(x',v')\int_0^{\tau_-(x,v)}\!\!\!\int_{\S^{n-1}}k(x-tv,v_1,v)
\int_0^{\tau_-(x-tv,v_1)}{E(x,x-tv,x-tv-t_1v_1,x')\over |x-tv-t_1v_1-x'|^{n-1}}\nonumber\\
&&\hspace{-2cm}\times k(x-tv-t_1v_1,v_2,v_1)k(x',v',v_2)_{v_2=\widehat{x-tv-t_1v_1-x'}}dt_1dv_1dtdx'dv',\label{t3}
\end{eqnarray}
for a.e. $(x,v)\in X\times\S^{n-1}$ and for $\psi\in L^1(X\times\S^{n-1})$. 
Therefore performing the change of variables $z=x-tv-t_1v_1$ ($dz=t_1^{n-1}dt_1dv_1$) on the right hand side of \eqref{t2} we obtain
\begin{eqnarray}
&&\beta_3(x,v,x',v')=\int_0^{\tau_-(x,v)}\int_X{E(x, x-tv,z,x')k(x',v',\widehat{z-x'})\over |x-tv-z|^{n-1}|z-x'|^{n-1}}\nonumber\\
&&\times k(x-tv,\widehat{x-tv-z},v)k(z,\widehat{z-x'},\widehat{x-tv-z})dtdz,\label{t5}
\end{eqnarray}
for a.e. $(x,v,x',v')\in X\times\S^{n-1}\times X\times \S^{n-1}$.
Then by induction we have
\begin{eqnarray}
\hspace{-2cm}
&&\beta_m(x,v,z_m,v_m)=\int_0^{\tau_-(x,v)}\int_{X^{m-2}}{E(x,x-tv,z_2,\ldots, z_m)\over |x-tv-z_2|^{n-1}\Pi_{i=2}^{m-1}|z_i-z_{i+1}|^{n-1}}\label{t6}\\
\hspace{-2cm}
&&\times k(x-tv,v_1,v)\Pi_{i=2}^mk(z_i,v_i,v_{i-1})_{|v_1=\widehat{x-tv-z_2},\ v_i=\widehat{z_i-z_{i+1}},\ i=2\ldots m-1}\nonumber
  dtdz_2\ldots dz_{m-1},\nonumber
\end{eqnarray}
for a.e. $(x,v,x',v')\in X\times\S^{n-1}\times X\times\S^{n-1}$.

We prove \eqref{eq:multiple} for $m=2$. From \eqref{t2},
\eqref{eq:barK} and \eqref{eq:J} it follows that
\begin{eqnarray}
  \hspace{-2.5cm}&&\bar {K^2}J\psi(x)=\int_{\S^{n-1}}\hspace{-.5cm}
  \sigma_a(x,v)\int_{X\times\S^{n-1}}\int_0^{\tau_-(x,v)}\hspace{-1cm}
  E(x,x-tv,y)
  {k(x-tv,v_1,v)k(y,v',v_1)_{|v_1= \widehat{x-tv-y}}\over 
     |x-tv-y|^{n-1}}dt\nonumber\\
  \hspace{-2.5cm}&&\times E(y,y-\tau_-(y,v')v')
  \psi(y-\tau_-(y,v')v',v')dydv'dv,\label{t20}
\end{eqnarray}
for a.e. $x\in X$ and for $\psi\in L^1(\Gamma_-,d\xi)$.  Then
performing the change of variable $z=x-tv$ ($t=|x-z|$,
$v=\widehat{x-z}$ and $dz=t^{n-1}dtdv$) on the right hand side of
\eqref{t20}, we obtain
\begin{eqnarray}
\hspace{-.5cm}&&\bar {K^2}J\psi(x)=\int_{X\times X\times\S^{n-1}}
  \hspace{-1cm}\sigma_a(x,\widehat{x-z})E(x,z,y)
{k(z,v_1,v)k(y,v',v_1)_{|v_1= \widehat{z-y}}\over |x-z|^{n-1}|z-y|^{n-1}}
\nonumber\\
\hspace{-.5cm}
&&\times E(y,y-\tau_-(y,v')v')\psi(y-\tau_-(y,v')v',v')dzdydv',\label{t21}
\end{eqnarray}
for a.e. $x\in X$ and for $\psi\in L^1(\Gamma_-,d\xi)$.  Performing
the change of variables $y=x'+t'v'$ ($x'\in \pa X$, $t'>0$,
$dz=|\nu(x')\cdot v'|d\mu(x')dt'$) on the right hand side of
\eqref{t21}, we obtain $\bar
{K^2}J\psi(x)=\int_{\Gamma_-}\alpha_2(x,x',v')\psi(x',v')d\mu(x')dv'$
for a.e. $x\in X$ and for $\psi\in L^1(\Gamma_-,d\xi)$, which proves
\eqref{eq:multiple} for $m=2$.

Then we prove \eqref{eq:gamma} before proving \eqref{eq:multiple} for
$m\ge 3$.  Let $m\ge 3$. From \eqref{t6} and the definition of the
operator $\bar {K^m}$ \eqref{eq:barK}, it follows that
\begin{eqnarray}
&&\bar {K^m}\psi(z_0)=\int_{\S^{n-1}}\sigma_a(z_0,v_0)\int_{X\times \S^{n-1}}\beta_m(z_0,v_0,z_m,v_m)\psi(z_m,v_m)dz_mdv_mdv_0\nonumber\\
&&=\int_{\S^{n-1}}\sigma_a(z_0,v_0)\int_{X\times\S^{n-1}}\int_0^{\tau_-(z_0,v_0)}\int_{X^{m-2}}E(z_0,z_0-tv_0,z_2,\ldots, z_m)\nonumber\\
&&\times {k(z_0-tv_0,v_1,v_0)\Pi_{i=2}^mk(z_i,v_i,v_{i-1})_{|v_1=\widehat{z_0-tv_0-z_2},\ v_i=\widehat{z_i-z_{i+1}},\ i=2\ldots m-1}\over |z_0-tv_0-z_2|^{n-1}\Pi_{i=2}^{m-1}|z_i-z_{i+1}|^{n-1}}
\nonumber\\
&&\times dtdz_2\ldots dz_{m-1}\psi(z_m,v_m)dz_mdv_mdv_0,\label{t7}
\end{eqnarray}
for a.e. $z_0\in X$ and for $\psi\in L^1(X\times\S^{n-1})$.  Therefore
performing the change of variables $z_1=z_0-tv_0$ ($dz_1=t^{n-1}dt
dv_0$, $t=|z_0-z_1|$ and $v_0=\widehat{z_0-z_1}$) on the right hand
side of \eqref{t7} we obtain $\bar
{K^m}\psi(z_0)=\int_{X\times\S^{n-1}}\gamma_m(z_0,z_m,v_m)\psi(z_m,v_m)$
$dz_mdv_m$, for a.e. $z_0\in X$ and for $\psi\in
L^1(X\times\S^{n-1})$, which proves \eqref{eq:gamma}.

We prove \eqref{eq:multiple}.  Let $m\ge 3$. From \eqref{eq:gamma},
\eqref{eq:barK} and \eqref{eq:J}, it follows that
\begin{eqnarray}
\hspace{-2cm}
&&\bar {K^m}J\psi(z_0)=\int_{X\times\S^{n-1}}\gamma_m(z_0,z_m,v_m)\nonumber\\
\hspace{-2cm}&&\times E(z_m,z_m-\tau_-(z_m,v_m)v_m)
\psi(z_m-\tau_-(z_m,v_m)v_m,v_m)dz_mdv_m\nonumber\\
\hspace{-2cm}
&&=\int_{X\times\S^{n-1}}\int_{X^{m-1}}
{E(z_0,\ldots,z_m,z_m-\tau_-(z_m,v_m)v_m)
\over\Pi_{i=1}^m|z_i-z_{i-1}|^{n-1}}\psi(z_m-\tau_-(z_m,v_m)v_m,v_m)
\label{t30}\\
\hspace{-2cm}
&&\times \left[\sigma_a(z_0,v_0)\Pi_{i=1}^mk(z_i,v_i,v_{i-1})
\right]_{v_i=\widehat{z_i-z_{i+1}},\ i=0\ldots m-1}
dz_1\ldots dz_{m-1}dz_mdv_m,\nonumber
\end{eqnarray}
for a.e. $z_0\in X$ and for $\psi\in L^1(\Gamma_-,d\xi)$.  Then
performing the change of variables $z_m=$ ``$z_m$'' $+t'v_m$
(``$z_m$''$\in \pa X$, $t'>0$, $dz_m=|\nu($``$z_m$''$)\cdot
v_m|d\mu($``$z_m$''$)dt'$), we obtain $\bar
{K^m}J\psi(z_0)=\int_{\Gamma_-}\alpha_m(z_0,z_m,v_m)\psi(z_m,v_m)d\mu(z_m)dv_m$
for a.e. $z_0\in X$ and for $\psi\in L^1(\Gamma_-,d\xi)$, which proves
\eqref{eq:multiple}.$\square$

\section*{Acknowledgment} 
This work was supported in part by NSF Grants DMS-0554097 and
DMS-0804696.



\end{document}